\pdfoutput=1 

\documentclass[final,3p,times,onecolumn]{elsarticle}

\usepackage[linesnumbered,ruled]{algorithm2e}

\usepackage{algpseudocode}
\usepackage{multirow,bigstrut}
\usepackage{enumitem}
\usepackage{arydshln}
\usepackage{empheq}
\usepackage{datetime}
\usepackage{bm}
\usepackage{threeparttable}

\newlength\figsep
\newlength\OneImW
\newlength\TwoImW
\newlength\FourImW
\newlength\BigOneImW
\newlength\Onefigwidth
\usepackage{graphicx,color}
\graphicspath{{figures-pdf/}}
\usepackage{enumitem}
\usepackage{subfigure}
\usepackage{footnote}

\journal{Elsevier}
\usepackage{hyperref}

\newtheorem{Property}{Property}
\newtheorem{Remark}{Remark}
\newtheorem{proof}{Proof}

\begin{document}

\begin{frontmatter}

\title{Security Measurement of a Medical Communication Scheme based on Chaos and DNA Coding}

\author[cn-xtu]{Lei Chen}

\author[cn-xtu]{Chengqing Li\corref{cor1}}

\author[cn-hnu]{Chao Li}

\cortext[cor1]{Corresponding author.}


\address[cn-xtu]{School of Computer Science, Xiangtan University, Xiangtan 411105, Hunan, China}

\address[cn-hnu]{College of Computer Science and Electronic Engineering, Hunan University, Changsha 410082, Hunan, China}

\begin{abstract}
To encrypt sensitive information existing in a color DICOM images, a medical privacy protection scheme (called as MPPS) based on chaos and DNA coding was proposed by
using two coupled chaotic systems to produce cryptographic primitives. Relying on some empirical analyses and experimental results, the designers of MPPS claimed that it
can withstand a chosen-plaintext attack and some other classic attacking models. However, this statement is groundless.
In this paper, we investigate the essential properties of MPPS and DNA coding, and we then propose an efficient chosen-plaintext attack to disclose its equivalent secret-key.
The attack only needs $\lceil \log_{256}(3\cdot M\cdot N)\rceil+4$ pair of chosen plain-images and the corresponding cipher-images, where $M \times N$ and ``3" are
the size of the RGB color image and the number of color channels, respectively. In addition, the other claimed superiorities are questioned from the perspective of modern cryptography.
Both theoretical and experimental results are presented to support the efficiency of the proposed attack and the other reported security faults.
The proposed cryptanalysis results will promote the proper application of DNA encoding to protect multimedia privacy data, especially that in a DICOM image.
\end{abstract}
\end{frontmatter}

\section{Introduction}

DNA (Deoxyribonucleic acid) computing is an emerging new generation of computing technology that uses biochemical reaction techniques to perform calculation in biological DNA molecules as information carriers
rather than the standard artificial hardware \cite{adleman1994molecular}.
DNA computing has some potential advantages over conventional alternatives, such as huge storage space, massive parallelism, and ultra-low power consumption \cite{lipton1995DNA}.
Consequently, it has received attention from many researchers in multiple fields, such as mathematics, biology, and computer science \cite{braich2002solution, clelland1999hiding}.
Some researchers have adopted the concept of DNA computing into cryptography, and adopted its computing rules as part of an encryption algorithm.
These algorithms are known as DNA cryptography \cite{leier2000cryptography}.
In 2004, Gehani et al. proposed the first DNA-based cryptosystem in \cite{Gehani:DNA:LNCS04}, which combines a one-time-pad and the diffusion operation based on exclusive OR operation.
A number of DNA-based encryption schemes have since been published in the literature and in patents.
Most of these schemes have adopted nonlinear dynamics for better encryption effect, including 2-D Logistic-adjusted-Sine map \cite{Wuxj:DNA:ND2017}, H\'{e}non-Sine map \cite{Liao:DNA:SP2018} and discrete memristor \cite{peng:meri:AEU21}.

Cryptography and cryptanalysis are two equivalent parts of cryptology. As shown in the development history of the Data Encryption Standard (DES), they can promote development of the two parts.
Some encryption schemes based on DNA encoding were found to be insecure to different extents from the viewpoint of modern cryptology
\cite{liu:DNA:optik2014,Hermassi:analysis:MTA2014,Su:analysis:MTA2017,Akhavan:DNA:OLT2017,Panwar:DNA:IJBC2019,Wangy:analysis:CEE2015,Wen:DNA:En2019}.
For example, Xie et al. evaluated the security performance of an image encryption algorithm based on DNA sequence operation and a hyper-chaotic system. They disclosed its equivalent
secret key with no more than $\lceil \log_2(4\cdot M \cdot N)/2 \rceil+1$ chosen plain-images, where $M\times N$ is the size of the plain-image \cite{liu:DNA:optik2014}.
In 2014, Hermassi et al. cryptanalyzed an image encryption algorithm based on DNA addition, and recovered its keystream (i.e., the equivalent secret-key) with some pairs
of chosen plain-images and the corresponding cipher-images \cite{Hermassi:analysis:MTA2014}.
From a survey of the literature it can be seen that most efforts have focused on using chosen-plaintext attacks as powerful attacking tools \cite{Su:analysis:MTA2017,Akhavan:DNA:OLT2017,Panwar:DNA:IJBC2019},
while a few works have proposed specific attack algorithms under the conditions of known-plaintext attacks \cite{Wangy:analysis:CEE2015} and chosen-ciphertext attacks \cite{Wen:DNA:En2019}.
Using the properties of DNA coding during encryption process, some cryptanalysis works have presented near-optimal attacking methods \cite{Belazi:analysis:ND2014, Zhangyong:Optick15,chen2020cryptanalysis}.
Specifically, DNA addition is modelled as equivalent version in terms of a binary addition in \cite{Belazi:analysis:ND2014};
a series of DNA manipulations can be viewed as S-box components in \cite{chen2020cryptanalysis}; and different DNA coding and decoding rules have the same encryption effect
in \cite{Zhangyong:Optick15}.

The fast development of transmission and sensing technologies supports the progress of remote diagnosis (telediagnosis) and remote surgery (telesurgery)\cite{chen:csf:TII2021}.
However, concerns about the security and privacy of the medical images transmitted via public channels are becoming increasingly serious\cite{chen:csencryption:opti2018}.
Due to the bulky size of medical image data and the special storage format, the modern text encryption standards are not efficient to protect them \cite{acharya:compact:ITB01,Zhou:Synch:SJ20}.
The directly optimized image protection strategy is to reduce the size of encryption object by automatically
detecting the region of interest (ROI) and leaving the region of non-interest (RONI) in plain-form.
As the equivalent counterpart of cryptography, the object of cryptanalysis is to obtain as much information on the secret-key and/or the plaintext as possible under a given attacking scenario  \cite{Ozkaynak:analysis:SIU2013,chenlei:cryptanalysis:CBM15,Zhucongxu:Tent:Access18,chencryptanalysis:ND18,wangsh:analysis:IS21}.
Among them, reference \cite{chenlei:cryptanalysis:CBM15} first eliminated the diffusion effect by using a differential attacking approach and then revealed the equivalent secret-key of the permutation process; while
reference \cite{Zhucongxu:Tent:Access18} obtained the keystream and permutation matrix by an efficient chosen-plaintext attacking method.

The degree of randomness of the pseudo-random number sequences (PRNSs) generated by iterating a chaotic map
in a digital device may be weaker than that of the counterpart obtained in an infinite domain \cite{cqli:delay:IEEEM22,cqli:block:JISA20,cqli:Cat:TC22,peng:meri:AEU21}.
As comprehensively reviewed in \cite{cqli:network:TCASI2019}, various methods have been proposed to counteract the dynamics degradation of digitized chaotic maps, as follows:
selecting state and control parameters; increasing the arithmetic precision; perturbing the control parameters; perturbing states;
switching among multiple chaotic maps; cascading multiple chaotic maps together \cite{Hua:SP:2018,hua:Sine:TIE2018}; and constructing much more complex systems \cite{cqli:Diode:TCASI19}.
For a critical review of chaotic cryptology, we refer the reader to \cite{Alvarez:IJBC:2006,ozkaynak:review:ND18,cqli:meet:JISA19}.

A medical privacy protection scheme (MPPS) based on DNA encoding and chaotic maps
was proposed in \cite{Ravichandran:DNA:ITN2017}. The images stored as DICOM (Digital Imaging and Communications in Medicine) standard
are encrypted with DNA encoding and PRNSs generated by iterating two  coupled chaotic systems (CCSs).
Based on some experimental results and empirical analysis, the designers of MPPS claimed that the encryption scheme is secure against chosen-plaintext attacks.
However, rigorous modern cryptanalysis reveals that this statement is groundless. This paper focuses on security evaluation of MPPS.
Some properties on its essential structure are reported and proved to be rigorous, which is then used to support an efficient chosen-plaintext attack.
A divide-and-conquer strategy is adopted to obtain the secret key of MPPS and/or its equivalent.
The metrics on validating security performance of MPPS are critically checked with some convincing arguments.

In general, the primary contributions of this paper are summarized as follows:
\begin{enumerate}[label=\arabic*)]
	\item The mathematical properties of DNA coding are studied;
	\item A divide-and-conquer attack is proposed to crack the three channels of the cipher-image produced by MPPS (namely Red, Green, and Blue channels of the color image);
	\item Both theoretical and simulation results demonstrate that there are a large number of equivalent keys of MPPS;
	\item The security performance of MPPS is comprehensively evaluated from eight aspects.
\end{enumerate}

The rest of this paper is organized as follows. Section~\ref{algorithm} presents the operations of the medical privacy protection scheme under study.
A detailed description of cryptanalysis on MPPS is presented in Sec.~\ref{cryptanalysis}, together with some experimental results. Finally, the last section concludes the paper with final remarks.

\section{A precise and concise description of MPPS}
\label{algorithm}

As specified in \cite[Sec.~III]{Ravichandran:DNA:ITN2017}, MPPS can protect the sensitive information in a DICOM image with two different modes: partial and full encryption.
If partial encryption is adopted, then only the significant areas of the DICOM image are encrypted to considerably reduce the size of the encryption object for diagnosis.
Except the difference on selected operating portions of a plain-image, MPPS works exactly the same in the two modes.
Without loss of generality, the encryption object of MPPS can be simplified as a RGB colour image $\bm{I}=\{\bm{I}_r, \bm{I}_g, \bm{I}_b\}$ of size $M\times N$ for either mode,
which are represented with three two-dimensional (2D) 8-bit integer matrices:
$\bm{I}_r=\{I_r(i, j)\}_{i=1, j=1}^{M, N}$, $\bm{I}_g=\{I_g(i, j)\}_{i=1, j=1}^{M, N}$ and $\bm{I}_b=\{I_b(i, j)\}_{i=1, j=1}^{M, N}$.
Accordingly, $\bm{I}'=\{\bm{I}'_r, \bm{I}'_g, \bm{I}'_b\}$ is the cipher-image, where
$\bm{I}'_r=\{I'_r(i, j)\}_{i=1, j=1}^{M, N}$, $\bm{I}'_g=\{I'_g(i, j)\}_{i=1, j=1}^{M, N}$ and $\bm{I}'_b=\{I'_b(i, j)\}_{i=1, j=1}^{M, N}$.
Each 2-D image data can also be written as a one-dimensional (1D) array by scanning it in the raster order (from left to right, and top to bottom);
for example, $\bm{I}_r=\{I_r(i)\}_{i = 1}^{M \times N}$ \footnote{Because the 2-D and 1-D presentation forms can be mutually converted,
they are not distinguished strictly in the paper.}.

\begin{figure*}[!htb]
	\centering
	\includegraphics[scale=0.65]{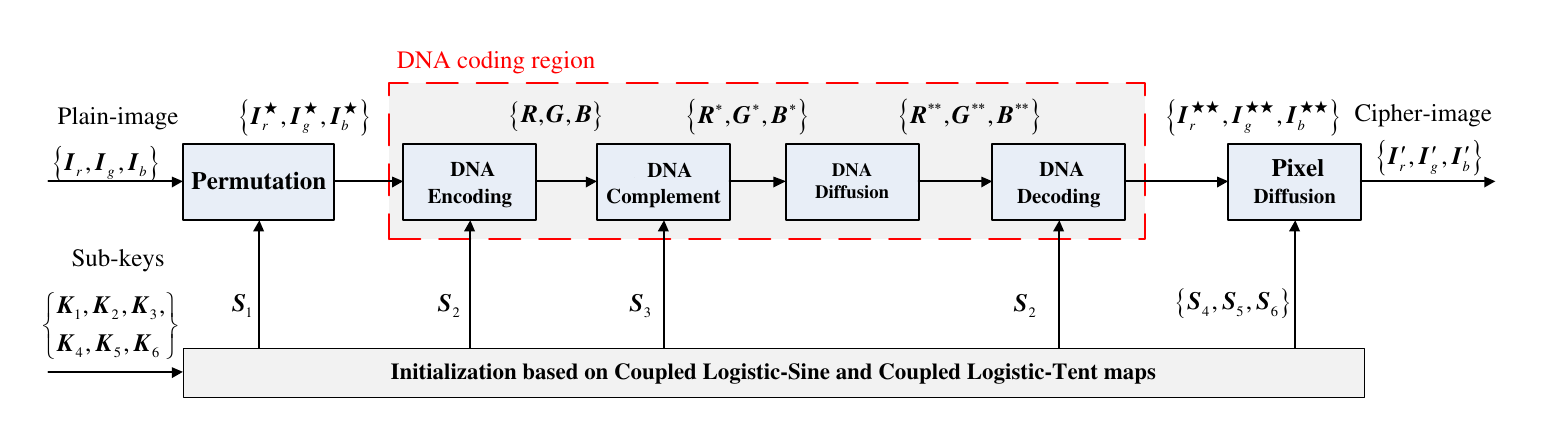}
	\caption{The overall architecture of MPPS.}
	\label{fig:dna_encryption_framework}
\end{figure*}

The architecture of MPPS designed in \cite{Ravichandran:DNA:ITN2017} is illustrated in Fig.~\ref{fig:dna_encryption_framework}.
The basic parts of MPPS are described as follows\footnote{To make the description of MPPS more concise and complete, some details of its process given in \cite{Ravichandran:DNA:ITN2017} are modified under the precondition that its security performance is uninfluenced.}:
\begin{itemize}[leftmargin=0.3cm]
\item \textit{The secret key} includes six sets of initial values and control parameters
$\{(Y_m(0), \mu_m)\}_{m=1}^6=\{K_m\}_{m=1}^6$. The sub-keys $K_1$, $K_4$, $K_5$ and $K_6$ are the initial values and control parameters of the  Coupled Logistic-Sine (CLS) map
\begin{equation}
\label{eq:cls}
f_{\textup{CLS}}(x)=f(x)+ \tfrac{1}{4}\cdot(4-\mu)\cdot\sin(\pi x) \bmod 1,
\end{equation}
where
\begin{equation}
\label{eq:logistic}
f(x)=\mu\cdot x\cdot(1-x)
\end{equation}
is the Logistic map studied in \cite{cqli:network:TCASI2019}. The other two sub-keys, $K_2$ and $K_3$, are sets of the initial values and control parameters of  a Coupled Logistic-Tent (CLT) map
\begin{equation}
\label{eq:cts}
f_{\textup{CLT}}(x)=
\begin{cases}
f(x)+\frac{1}{2}(4-\mu)\cdot x \bmod 1     & \mbox{if } x<0.5; \\
f(x)+\frac{1}{2}(4-\mu)\cdot (1-x) \bmod 1 & \mbox{if } x\geq 0.5,
\end{cases}
\end{equation}
where $x\in [0,1]$ and $\mu\in[0, 4]$.

\item\textit{Initialization}:

\begin{enumerate}[label=\arabic*)]
\item \textit{Generating an index sequence}:
iterate CLS map~(\ref{eq:cls}) $t_1$ times from the initial condition $Y_1(0)$ with control parameter $\mu_1$ to avoid the transient effect generated in the initial iterations.
Iterate it $3L$ more times to obtain a state sequence $\bm{X}_1=\{X_1(i)\}_{i=1}^{3L}$, where $L=M \times N$.
Sort $\bm{X}_1$ in an ascending order to produce an index sequence $\bm{S}_1=\{S_{1}(i)\}_{i=1}^{3L}$ satisfying that
$\bm{X}_1(S_1(i))$ is the $i$-th largest element of sequence $\bm{X}_1$.

\item \textit{Determining DNA coding rules}:
iterate CLT map (\ref{eq:cts}) with control parameter $\mu_2$ from initial condition $Y_2(0)$ $t_2$ times to avoid the transient effect.
Iterate it six more times to obtain a sequence $\bm{X}_2=\{X_{2}(i)\}_{i=1}^6$ and quantize it to six random numbers $\{S_2(i)\}_{i=1}^6$ via
\begin{equation}
S_2(i)= \lfloor X_{2}(i) \times 10^{14} \rfloor\bmod 8,
\label{eq:convert14}
\end{equation}
where $\lfloor x \rfloor$ returns the nearest integer that is less than or equal to $x$.

\item \textit{Constructing a pseudo-random number generator}:
starting from the initial condition $Y_3(0)$, iterate CLT map (\ref{eq:cts}) $t_3$ times with control parameter $\mu_3$.
Iterate map~(\ref{eq:cts}) $4L$ more times to obtain a state sequence $\bm{X}_3=\{X_{3}(i)\}_{i=1}^{4L}$ and quantize it to a binary sequence $\bm{S}_3=\{S_3(i)\}_{i=1}^{4L}$ via
\begin{equation}
\label{eq:bin_seq}
S_3(i)=
\begin{cases}
0  & \mbox{if } 0 \leq X_3(i) \leq 0.5;\\
1  & \mbox{if } 0.5<X_3(i) < 1.
\end{cases}
\end{equation}

\item \textit{Generating three keystreams $S_4$, $S_5$, $S_6$}:
similar to Step 1, set three sub-keys $K_4$, $K_5$, and $K_6$ as the control parameter and
initial state of CLS map~(\ref{eq:cls}), generate three sequences $\bm{X}_4 =\{X_4(i)\}_{i=1}^{L}$, $\bm{X}_5=\{X_{5}(i)\}_{i=1}^{L}$ and $\bm{X}_6 =\{X_6(i)\}_{i=1}^{L}$, respectively.
As above, the first $t_m$ states are eliminated to generate $S_m$, where $m=4, 5, 6$. Then, every element of these sequences is quantified via function
\begin{equation}
\label{eq:q_s4}
S_m(i)=\lfloor X_{m}(i)\cdot 10^{14} \rfloor \bmod 256.
\end{equation}
\end{enumerate}

\begin{table}[!htb]
	\centering
	\caption{Eight different DNA coding rules}
	\begin{tabular}{ccccc}
		\hline
		Rule number   &$(00)_2$  &$(01)_2$  &$(10)_2$  &$(11)_2$ \\ \hline
		$0$  &A   &C   &G   &T  \\
		$1$  &A   &G   &C   &T  \\
		$2$  &T   &G   &C   &A  \\
		$3$  &T   &C   &G   &A  \\
		$4$  &C   &A   &T   &G  \\
		$5$  &C   &T   &A   &G  \\
		$6$  &G   &A   &T   &C  \\
		$7$  &G   &T   &A   &C  \\ \hline
	\end{tabular}
\label{tab:dna_coding}
\end{table}

\item\textit{Encryption procedure}:
\begin{enumerate}[label=\alph*)]
\item \textit{Permutation}:	
three matrices $\bm{I}_r$, $\bm{I}_g$, and $\bm{I}_b$ are ``placed" vertically to obtain a larger image $\bm{P}=\{P(i, j)\}_{i=1, j=1}^{3M, N}$.
Then, produce a scrambled intermediate image $\bm{P}^{\star}=\{P^{\star}(i, j)\}_{i=1, j=1}^{3M, N}$ from $\bm{P}$ by the index sequence $\bm{S}_1$:
\begin{equation}
\label{eq:permut}
P^{*}(i, j)=P(u, v),
\end{equation}
where
\begin{equation}
\label{eq:uv}
\begin{cases}
i = \left\lfloor \tfrac{S_1((u-1)\cdot N+v-1)}{N} \right\rfloor+1; \\
j = S_{1}((u-1)\cdot N+v-1)\bmod N +1.
\end{cases}
\end{equation}

\item \textit{DNA Encoding}:
first, assign $E_i=X_2(i)$ for $i=1\sim 3$.
Using transform function
\begin{equation}
q(x)=(x_3, x_2, x_1, x_0)
\label{eq:2bits}
\end{equation}
satisfying $\sum_{i=0}^{3}x_i\cdot 2^{2i}=x$ to change every pixel of $\{P^{\star}(i, j)\}_{i=1, j=1}^{M, N}$, $\{P^{\star}(i, j)\}_{i=M+1, j=1}^{2M, N}$, and $\{P^{\star}(i, j)\}_{i=2M+1, j=1}^{3M, N}$ into four neighboring elements of 2-bit integer matrices of size $1\times 4L$, $\bm{I}^{\star}_r$, $\bm{I}^{\star}_g$, and $\bm{I}^{\star}_b$, respectively.
For example, the pixel "228" can be represented as $x = 228 = 3 \times 2^{6} + 2 \times 2^{4} + 1 \times 2^{2} + 0 \times 2^{0}$, thus  the 2-bit integer is $(3,2,1,0)$.
Then, convert $\bm{I}^{\star}_r$ , $\bm{I}^{\star}_g$ and $\bm{I}^{\star}_b$ into DNA symbol matrices with the encoding rules in the $E_1$-th, $E_2$-th, and $E_3$-th row of Table~\ref{tab:dna_coding}, respectively.
Let $\bm{R}$, $\bm{G}$ and $\bm{B}$ denote the corresponding conversion results.

\item \textit{DNA Complement}:
for Red channel, obtain matrix $\bm{R^*}=\{R^*(i)\}_{i=1}^{4L}$ via
\begin{equation}
\label{eq:dna_trans}
R^*(i)=
\begin{cases}
R(i)      & \mbox{if~} S_{3}(i)=0;\\
g(R(i))   & \mbox{if~} S_{3}(i)=1,
\end{cases}
\end{equation}
where
\begin{equation}
\label{eq:dna_gx}
g(x) =
\begin{cases}
T      & \mbox{if~} x=``A"; \\
C      & \mbox{if~} x=``T"; \\
G      & \mbox{if~} x=``C"; \\
A      & \mbox{if~} x=``G".
\end{cases}
\end{equation}
For Green and Blue channels, they remain unchanged, i.e., $\bm{G^*}= \bm{G}$, $\bm{B^*}= \bm{B}$.

\item \textit{DNA Diffusion}:
According to the DNA addition operation rules defined in Table~\ref{tab:dna_add}, calculate
\begin{equation}
\label{eq:dna_diffusion}
\begin{cases}
	R^{**}(i)   = R^{*}(i);\\
	G^{**}(i)   = G^{*}(i) \boxplus R^{*}(i);\\
	B^{**}(i)   = B^{*}(i) \boxplus G^{*}(i);
\end{cases}
\end{equation}
for $i=1, 2\sim 4L$.

\item \textit{DNA Decoding}:
first, assign $D_i=X_2(i+3)$ for $i=1, 2 3$.
Then, change DNA symbol matrices $\bm{R}^{**}$, $\bm{G}^{**}$, and $\bm{B}^{**}$ into 2-bit integer matrices with the conversion rule in the $D_1$-th, $D_2$-th, and $D_3$-th rows of Table~\ref{tab:dna_coding}, respectively.
Accordingly, intermediate matrices $\bm{R}^{\star \star}$, $\bm{G}^{\star \star}$, and $\bm{B}^{\star \star}$ are produced.
Finally, transform $\bm{R}^{\star \star}$, $\bm{G}^{\star \star}$, and $\bm{B}^{\star \star}$
into $\bm{I}^{\star \star}_r$, $\bm{I}^{\star \star}_g$, and $\bm{I}^{\star \star}_b$, respectively.
In these transforms, every four consecutive 2-bit integer elements are merged into one 8-bit integer with the inverse function of Eq.~(\ref{eq:2bits}).

\item \textit{Pixel Diffusion}:
considering every 2-D matrix as a 1-D sequence by scanning it in the raster order, produce matrices $\bm{I}'_r, \bm{I}'_g, \bm{I}'_b$ by calculating
\begin{equation}
\label{eq:pixel_diffusion}
\left\{
\begin{split}
I'_r(i) & = I'_r(i-1) \oplus {I}^{\star \star}_r(i) \oplus S_4(i);\\
I'_g(i) & = I'_g(i-1) \oplus {I}^{\star \star}_g(i) \oplus{I}^{\star \star}_b(i) \oplus S_{5}(i);\\
I'_b(i) & = I'_b(i-1) \oplus {I}^{\star \star}_b(i) \oplus{I}^{\star \star}_r(i) \oplus S_{6}(i),
\end{split}
\right.
\end{equation}
$i=1\sim L$, $\oplus$ denotes the bitwise exclusive OR operation, $I'_r(0)=0$, $I'_g(0)=0$, and ${I}'_b(0)=0$.
\end{enumerate}

\item \textit{Decryption procedure}:
the decryption procedure is the inverse version of the above encryption procedure.
The DNA subtraction rules are defined in Table~\ref{tab:dna_sub}.
\end{itemize}

\begin{table}[!htb]
	\centering	
	\caption{DNA addition rules.}
	\begin{tabular}{lllll}
		\hline
		$\boxplus$  &A  	&G   &C   &T  \\ \hline
		A  &A   &G   &C   &T  \\
		G  &G   &C   &T   &A  \\
		C  &C   &T   &A   &G  \\
		T  &T   &A   &G   &C  \\ \hline
	\end{tabular}\label{tab:dna_add}
\end{table}

\begin{table}[!htb]
\centering
	\caption{DNA subtraction rules.}
	\begin{tabular}{lllll}
		\hline
		$\boxminus$  &A   &G   &C   &T  \\ \hline
		A  &A   &T   &C   &G  \\
		G  &G   &A   &T   &C  \\
		C  &C   &G   &A   &T  \\
		T  &T   &C   &G   &A  \\ \hline
	\end{tabular}
\label{tab:dna_sub}
\end{table}

\section{Cryptanalysis of MPPS}
\label{cryptanalysis}

To objectively evaluate the security performance of MPPS, its essential properties are first investigated. This is then used to underpin an efficient chosen-plaintext attack on it.
Finally, the eight security performance metrics of MPPS are re-evaluated one by one.

\subsection{Some essential properties of MPPS}
\label{subsec:properties}

To aid the subsequent theoretical analysis, some of the properties of MPPS and the DNA coding are analyzed.

First, a property of the differences between the ciphertexts generated by MPPS is introduced. The difference between
two cipher-images $\bm{I}'_1$ and $\bm{I}'_2$ is defined as $\Delta \bm{I}' = \bm{I}'_{1} \oplus \bm{I}'_{2} = \{\Delta \bm{I}'_{r}, \Delta \bm{I}'_{g}, \Delta \bm{I}'_{b} \} $, where $\Delta \bm{{I}'_{r}} = \bm{I}'_{r1} \oplus \bm{I}' _{r2}= \{ I'_{r1}(i) \oplus I'_{r2}(i) \} _{i=1}^{L}$, $\Delta \bm{{I}'_g}=\{ I'_{g1}(i) \oplus I'_{g2}(i) \} _{i=1}^{L}$, $\Delta \bm{{I}'_b} = \{ I'_{b1}(i) \oplus I'_{b2}(i) \} _{i=1}^{L}$.
Accordingly, the difference between intermediate images $\bm{I}^{\star\star}_{1}$ and $\bm{I}^{\star\star}_2$ is defined as
$\Delta \bm{I}^{\star \star}=\bm{I}^{\star \star}_{1} \oplus \bm{I}^{\star \star}_2=\{\Delta \bm{I}^{\star\star}_{r}, \Delta \bm{I}^{\star\star}_{g}, \Delta \bm{I}^{\star \star}_{b} \}$, where
$\Delta \bm{I^{\star \star}_{r}} =\bm{I}^{\star \star}_{r1} \oplus \bm{I}^{\star \star} _{r2}= \{I^{\star \star}_{r1}(i)\oplus I^{\star \star}_{r2}(i) \} _{i=1}^{L}$,
$\Delta \bm{{I}^{\star \star}_g}=\{I^{\star \star}_{g1}(i)\oplus I^{\star\star}_{g2}(i)\}_{i=1}^{L}$, $\Delta \bm{{I}^{\star \star}_b}=\{I^{\star \star}_{b1}(i)\oplus I^{\star\star}_{b2}(i)\}_{i=1}^{L}$.

\begin{Property}
The differences $\Delta\bm{I}'_r$, $\Delta \bm{I}'_g$ and $\Delta\bm{I}'_b$ are unrelated with the secret keys (and their equivalents); that is, $K_4$ ($\bm{S}_4$), $K_5$ ($\bm{S}_5$), and $K_6$ ($\bm{S}_6$).
\label{prop:differ}
\end{Property}
\begin{proof}
By observing Eq.~(\ref{eq:pixel_diffusion}), one can get
	\begin{equation}	
	\begin{cases}
	\Delta {I}'_r(i)  =  \Delta {I}'_{r}(i-1) \oplus \Delta {I}^{\star \star}_r(i); \\
	\Delta {I}'_g(i)  =  \Delta {I}'_{g}(i-1) \oplus \Delta {I}^{\star \star}_g(i) \oplus \Delta {I}^{\star \star}_b(i); \\
	\Delta {I}'_b(i)  =  \Delta {I}'_{b}(i-1) \oplus \Delta {I}^{\star \star}_b(i) \oplus \Delta {I}^{\star \star}_r(i),
	\end{cases}
    \label{eq:delta_cipher}
	\end{equation}
and then the property is proved.
\end{proof}

\begin{Remark}
Referring to Property~\ref{prop:differ} and its proof, one can decrypt the difference of cipher-images
$\{\Delta \bm{I}'_r, \Delta \bm{I}'_g, \Delta \bm{I}'_b\}$ to obtain the difference of intermediate images $\{\Delta \bm{I}^{\star\star}_r, \Delta \bm{I}^{\star \star}_g, \Delta \bm{I}^{\star\star}_b\}$
even without any information on the secret or an equivalent key.
\label{remark:decry}
\end{Remark}

Let $f_s(x)$ represent the DNA encoding defined in Table~\ref{tab:dna_coding}, where $s$ denotes a rule number; that is, $s\in\mathbb Z_8=\{0, 1, 2, 3, 4, 5, 6, 7\}$.
Accordingly, let $f_{t}^{-1}(x)$ denote decoding transformation, $g(x)$ represents DNA complement transformation given in Eq.~(\ref{eq:dna_gx}),
where $x\in\mathbb{Z}_4=\{0, 1, 2, 3\}$ and $t$ denote a rule number.

To investigate the intrinsic properties of computing of DNA coding, two composite functions $F_{s, t}(x)=f_{t}^{-1}(f_{s}(x))$ and $G_{s, t}(x)=f_{t}^{-1}(g(f_{s}(x)))$ are defined.

\begin{Property}
If $x_0\oplus x_1=(11)_2=3$, then one has
	\begin{equation}
	\label{eq:dna_relation}
	F_{s,t}(x_0) \oplus F_{s,t}(x_1) \neq G_{s,t}(x_0) \oplus G_{s,t}(x_1).
	\end{equation}	
where $x_0, x_1 \in\mathbb{Z}_4$ and $s, t\in\mathbb{Z}_8$.
\label{prop:dna_relation}
\end{Property}
\begin{proof}
Given that $x_0\oplus x_1=(11)_2$, there is a DNA complement between $f_{s}(x_1)$ and $f_{s}(x_2)$; that is,
\begin{align*}
(f_s(x_0), f_s(x_1))\in \{(A, T), (T, A), (C, G), (G, C)\}.
\end{align*}
By substituting the above result into Eq.~(\ref{eq:dna_gx}), one can get
\begin{align*}
(g(f_s(x_0)), g(f_s(x_1))) \in \{(T, C), (C, T), (G, A), (A, G)\}.
\end{align*}
By decoding $(f_s(x_0), f_s(x_1))$ and $(g(f_s(x_0)), g(f_s(x_1)))$ to obtain
$(f_t^{-1}(f_s(x_0)), f_t^{-1}(f_s(x_1)))$ (or $(F_{s, t}(x_0), F_{s, t}(x_1))$)
and $(f_t^{-1}(g(f_s(x_0)))$, $f_t^{-1}(g(f_s(x_1))))$ (or $(G_{s, t}(x_0), G_{s, t}(x_1))$),
one has
\begin{equation}
\label{eq:dna_relation_proof}
	\begin{cases}
	F_{s, t}(x_0) \oplus F_{s,t}(x_1) = (11)_2;\\
	G_{s, t}(x_0) \oplus G_{s,t}(x_1) \in \{(10)_2, (01)_2\}.
	\end{cases}
\end{equation}
Hence, $F_{s, t}(x_0)\oplus F_{s, t}(x_1)\neq G_{s, t}(x_0)\oplus G_{s, t}(x_1)$.
\end{proof}

\begin{Property}
Composite functions $F_{s, t}(x)$ and $G_{s, t}(x)$ are both bidirectional maps that are defined in domain $\mathbb{Z}_4$.
They only have eight and 16 different maps, respectively.
\label{prop:dna_mapping_num}
\end{Property}
\begin{proof}
First, $f_s(x)$, $f_t^{-1}(x)$ and $g(x)$ are all bidirectional maps that are defined in domain $\mathbb{Z}_4$,
so $F_{s, t}(x)=f_t^{-1}(f_{s}(x))$ and $G_{s, t}(x)=f_t^{-1}(g(f_s(x)))$ are also fixed bidirectional maps.	
Considering the constraint of Eq.~(\ref{eq:dna_relation_proof}), one can enumerate that the possible numbers of $F_{s, t}(x)$ and $G_{s, t}(x)$ are
$\binom{4}{1}\cdot\binom{2}{1}=8$
and
$\binom{4}{1}\cdot\binom{2}{1}\cdot\binom{2}{1}=16$,
respectively.
\end{proof}

\begin{Property}
Given a map $F_{s_1, t_1}(x)$, there are different versions of $F_{s_2, t_2}(x)$ satisfying
	\begin{equation}	
	F_{s_1, t_1}(x)\oplus F_{s_2, t_2}(x)=\lambda
    \label{eq:fx_xor_keys}
	\end{equation}
for any $x$, where $\lambda$ is a fixed value and $\lambda\in\{1, 2, 3\}$,
and $x\in\mathbb{Z}_4$, $s_1, s_2, t_1, t_2\in\mathbb{Z}_8$.	
\label{prop:fx_xor_keys}	
\end{Property}
\begin{proof}
Referring to Eq.~(\ref{eq:dna_relation_proof}), one can get
\begin{align}
F_{s_1, t_1}(x_0)\oplus F_{s_1, t_1}(x_1) &= F_{s_1, t_1}(x_2)\oplus F_{s_1, t_1}(x_3)\nonumber\\
                                          &= F_{s_2, t_2}(x_0)\oplus F_{s_2, t_2}(x_1)\nonumber\\
	                                      &= F_{s_2, t_2}(x_2) \oplus F_{s_2,t_2}(x_3)\nonumber\\
	                                      &= (11)_2,
\end{align}
where $x_0\oplus x_1=x_2\oplus x_3=(11)_2$, $x_0\neq x_1\neq x_2\neq x_3$, and $x_0, x_1, x_2, x_3\in\mathbb{Z}_4$.
Then, one has
\begin{align}
F_{s_1, t_1}(x_0) \oplus F_{s_2, t_2}(x_0)&= F_{s_1, t_1}(x_1) \oplus F_{s_2, t_2}(x_1)\nonumber \\
                                          &= F_{s_1, t_1}(x_2) \oplus F_{s_2, t_2}(x_2)\nonumber \\
                                          &= F_{s_1, t_1}(x_3) \oplus F_{s_2, t_2}(x_3)\nonumber \\
                                          &= \lambda.
\end{align}
Given that $F_{s_1, t_1}(x)$ and $F_{s_2, t_2}(x)$ are two different maps, $\lambda\neq 0$, namely $\lambda \in \{1, 2, 3\}$.
\end{proof}

\begin{Property}
Given a map $G_{s_1, t_1}(x)$, there are different versions of map $G_{s_2, t_2}(x)$ satisfying
\begin{equation}
G_{s_1, t_1}(x) \oplus G_{s_2, t_2}(x)=\lambda
\label{eq:gx_xor_keys}
\end{equation}
for any $x$, where $\lambda$ is a fixed value and $\lambda\in\{1, 2, 3\}$, and $x\in\mathbb{Z}_4$,
$s_1, s_2, t_1, t_2\in\mathbb{Z}_8$.
\label{prop:gx_xor_keys}
\end{Property}
\begin{proof}
Referring to Eq.~(\ref{eq:dna_relation_proof}), there are two possible cases:
\begin{itemize}
\item when $G_{s1, t1}(x_0)\oplus G_{s1, t1}(x_1)=G_{s2, t2}(x_0)\oplus G_{s2, t2}(x_1)=(10)_2$, $x_0 \oplus x_1 = (11)_2$;	

\item when $G_{s1, t1}(x_0) \oplus G_{s1,t1}(x_1) = G_{s2, t2}(x_0) \oplus G_{s2,t2}(x_1) = (01)_2$, $x_0 \oplus x_1 = (11)_2$.
\end{itemize}	
Similar to the proof of Property~\ref{prop:fx_xor_keys}, either case can be proven.
\end{proof}

\begin{table}[!htb]
\centering
\caption{The original sub-key of MPPS and the corresponding equivalent key.}
\begin{tabular}{ccc}
		\hline
		Original sub-key 	&Corresponding equivalent sub-key                      & Description                    \\ \hline
		${K}_1$        	&$\bm{S}_1$                                                & Permutation index sequence     \\
		${K}_2$        	&$\bm{S}_2$ or \newline{$(E_1, E_2, E_3, D_1, D_2, D_3)$}  & DNA coding and decoding rules  \\
		${K}_3$        	&$\bm{S}_3$ 									           & Binary sequence                \\
		${K}_4$        	&$\bm{S}_4$										           & Keystream of R plane           \\
		${K}_5$        	&$\bm{S}_5$										           & Keystream of G plane           \\
		${K}_6$		   	&$\bm{S}_6$										           & Keystream of B plane           \\ \hline		
\end{tabular}
\label{tab:equ_keys}
\end{table}

\subsection{Cryptoanalyzing MPPS with a chosen-plaintext attack}

Under the scenario of the chosen-plaintext attack, an attacker can disclose the equivalent secret key of MPPS with some chosen plaintexts.
From \textit{encryption procedure} a) in Sec.~\ref{algorithm}, one can see that the same permutation operation controlled by $K_1$ can be performed if $\bm{S}_1$ is available;
namely, $\bm{S}_1$ is the equivalent counterpart of $K_1$.
In addition, $K_2$, $K_3$, $K_4$, $K_5$, and $K_6$ have the corresponding equivalent sub-keys, which are presented in Table~\ref {tab:equ_keys}.

To aid the subsequent analysis, a set of special plain-images is defined: $\mathcal{I}=\{\bm{I}_0, \bm{I}_1, \ldots, \bm{I}_{255}\}=\{\{\bm{I}_{r0}, \bm{I}_{g0}, \bm{I}_{b0}\}$,
$\{\bm{I}_{r1}, \bm{I}_{g1}, \bm{I}_{b1}\}$, $\ldots, \{\bm{I}_{r255}, \bm{I}_{g255}, \bm{I}_{b255}\}\}$,
where
$\bm{I}_{r0}=\bm{I}_{g0}=\bm{I}_{b0}=\{0, 0, \ldots, 0\}, \bm{I}_{r1}=\bm{I}_{g1}=\bm{I}_{b1}=\{1, 1, \ldots, 1\}$
and
$\bm{I}_{r255}=\bm{I}_{g255}=\bm{I}_{b255}=\{255, 255, \ldots, 255\}$.
Assume that the corresponding cipher-image set is $\mathcal{I'}=\{\bm{I}'_0, \bm{I}'_1, \ldots, \bm{I}'_{255}\}$.
The plain-images are specially chosen to assure their permuted versions remain unchanged, namely
${\bm{I}}^{\star}_{0}=\bm {I}_{0}$,
${\bm{I}}^{\star}_{1}=\bm {I}_{1}, \ldots$,
${\bm{I}}^{\star}_{255}=\bm {I}_{255}$.
Thus, an attacker only needs to focus on the intermediate states from the permutated images to the cipher-images.
Recalling the description of MPPS, the DNA coding and diffusion operations after the permutation process are executed in the Red, Green, and Blue channels, separately.
To disclose the sub-key corresponding to each channel, a  divide-and-conquer (DAC) strategy is adopted in the further cryptanalysis.

\subsubsection{Cracking encryption operations in the Red channel}

By observing the change rule of the data in the red channel during the whole encryption process, one can reveal the equivalent sub-keys $\bm{S}_3$, $E_1$, $D_1$, and $\bm{S}_4$.
Choose two special images $\bm{I}_{\alpha}$, $\bm{I}_{\beta}$ from image set $\mathcal{I}$ satisfying $\bm{I}_{\alpha}\oplus\bm{I}_{\beta}=\bm{I}_{255}$, and $\alpha, \beta\in\{0, 1, 2, \cdots, 255\}$;
that is, images $\bm{I}_0=(00000000)_2$ and $\bm{I}_{255}=(11111111)_2$; or $\bm{I}_{85}=(01010101)_2$ and $\bm{I}_{170}=(10101010)_2$.
Obviously, there are $\frac{256}{2}=128$ possible image pairs $(\bm{I}_{\alpha}, \bm{I}_{\beta})$ satisfying the constraint.
As discussed earlier, the permutated versions are $\bm{I}^{\star}_{\alpha}=\bm{I}_{\alpha}=\{\alpha, \alpha, \cdots, \alpha\}$ and $\bm{I}^{\star}_{\beta}=\bm{I}_{\beta}=\{\beta, \beta, \cdots, \beta\}$.
Now, only the Red channel of the permuted images---that is, $\bm{I}^{\star}_{r\alpha}$ and $\bm{I}^{\star}_{r\beta}$---is left.
To describe the subsequent encryption process, take $\bm{I}^{\star}_{r \alpha}$ and its $i$-th pixel $I^{\star}_{r\alpha}(i)$ of value $\alpha$ as an example.
Using Eq.~(\ref{eq:2bits}), pixel $I^{\star}_{r\alpha}(i)=\alpha$ is encoded with four DNA codes:
\begin{equation}
\label{eq:real_fx}
(f_s(\alpha_3), f_s(\alpha_2), f_s(\alpha_1), f_s(\alpha_0)),
\end{equation}
where $\sum_{n=0}^{3}\alpha_n\cdot 2^{2n}=\alpha$.
Then, the earlier four DNA codes are complemented via Eq.~\eqref{eq:dna_trans} and Eq.~\eqref{eq:dna_gx}, namely
\begin{equation}
\label{eq:reveal_gx}
{R}^{*}_{\alpha}(l)=
\begin{cases}
f_{s}(\alpha_n)         & \mbox{if~} S_{3}(l)=0;\\
g(f_{s}(\alpha_n))      & \mbox{if~} S_{3}(l)=1,
\end{cases}
\end{equation}
where $l=4i-4+n+1$, $n\in \mathbb{Z}_4$, $i\in\mathbb{Z}_L$, and $l\in\mathbb{Z}_{4L}$.
Referring to the DNA diffusion in Eq.~(\ref{eq:dna_diffusion}), one can see that the Red channel  remains unchanged.
Thus, the four DNA codes in Eq.~(\ref{eq:reveal_gx}) can be decoded as
\begin{equation}
\label{eq:reveal_fx_1}
{R}^{**}_{\alpha}(l) =
\begin{cases}
f_t^{-1}(f_s(\alpha_n))    & \mbox{if~} S_3(l)=0;\\
f_t^{-1}(g(f_s(\alpha_n))) & \mbox{if~} S_3(l)=1.
\end{cases}
\end{equation}
Similarly, by substituting the $k$-th pixel ${I}^{\star}_{r \beta}(k)$ of value $\beta$ in the image $\bm{I}^{\star}_{r \beta}$ into Eqs.~(\ref{eq:real_fx}), (\ref{eq:reveal_gx}), and (\ref{eq:reveal_fx_1}), one
can get
\begin{equation}
 R^{**}_{\beta}(l) =
\label{eq:reveal_fx_1_beta}
\begin{cases}
f_t^{-1}(f_{s}(\beta_n))         & \mbox{if~} S_3(l)=0;\\
f_t^{-1}(g(f_{s}(\beta_n)))      & \mbox{if~} S_3(l)=1.
\end{cases}
\end{equation}
Given that $\alpha\oplus\beta=255$ and $\alpha_n\oplus\beta_n=(11)_2$, one has
\begin{equation}
S_{3}(l) =
\label{eq:reveal_s3}
\begin{cases}
0  & \mbox{if~} \Delta{R}^{**}(l)=(11)_2;\\
1  & \mbox{if~} \Delta{R}^{**}(l)\in\{(10)_2, (01)_2\},
\end{cases}
\end{equation}
from Property~\ref{prop:dna_relation} and Eq.~(\ref{eq:dna_relation_proof}), where $\Delta{R}^{**}(l)=R^{**}_{\alpha}(l)\oplus{R}^{**}_{\beta}(l)$.
This deduction is a forward analysis of two images $\bm{I}_{\alpha}$ and $\bm{I}_{\beta}$.
A backward analysis on them can then be considered by calculating the difference on the red channel of the corresponding cipher-images:
$\Delta\bm{I}'_r=\bm{I}'_{r\alpha}\oplus\bm{I}'_{r\beta}=\{I_{r\alpha}(i)\oplus I_{r\beta}(i)\}_{i=1}^L$.
Referring to Property~\ref{prop:differ} and Remark~\ref{remark:decry}, one can reversibly recover $\Delta\bm{I}^{\star\star}_r$ from $\Delta\bm{I}'_r$.
Consequently, every 2-bit integer $\Delta R^{**}(l)$ in $\Delta I^{\star \star}_r(k)$ is accessible to the attacker, and binary sequence $\bm{S}_3$ can be recovered
via Eq.~(\ref{eq:reveal_s3}).

The next task is to obtain $E_1$, $D_1$, and $\bm S_4$. Because the number of possible combination of $D_1$ and $E_1$ is only $8\times 8=64$, they can be exhaustively searched and verified
with the method illustrated in Algorithm~\ref{alg:keysearch}.
In the process of verifying the candidate keys, only four pairs of special plain-image and the corresponding cipher-images are needed: $(\bm{I}_0, \bm {I}'_0), (\bm{I}_{85}, \bm {I}'_{85}),
(\bm{I}_{170}, \bm{I}'_{170})$, and $(\bm{I}_{255}, \bm{I}'_{255})$, which correspond to four different plain pixels $(00000000)_2$, $(01010101)_2$, $(10101010)_2$, and $(11111111)_2$, respectively.
This is attributed to the fact that functions $F_{E_1, D_1}(x)$ and $G_{E_1, D_1}(x)$ only have four different inputs; that is, $x\in\{(00)_2, (01)_2, (10)_2, (11)_2\}$. Based on such analysis, it could also be four other plain-image, for example, and their pixel values correspond to $(00110000)_2$, $(01100101)_2$, $(10011010)_2$ and $(11001111)_2$ respectively.
In addition, from Eq.~(\ref{eq:pixel_diffusion}), one can decrypt the red channel of cipher-image $\bm{I}'$, $\bm{I}'_{r}$,
to obtain the intermediate image $\bm{I}^{\star\star}_r\oplus \bm S_4$ by
\begin{equation}
\label{eq:cipher_r_decry}
I^{\star \star}_r(i)\oplus S_4(i)=I'_r(i)\oplus I'_r(i-1).
\end{equation}
In general, the main idea of Algorithm~\ref{alg:keysearch} is that if $E_1$ and $D_1$ are both guessed correctly.
Then, candidate sequences $\bm S_4$ obtained from multiple pairs of plain-images and the corresponding cipher-images are consistent.

\begin{algorithm}[!htb]
\caption{The procedure of searching for the keys exhaustively.}
\label{alg:keysearch}
\DontPrintSemicolon
\SetKwProg{Fn}{Function}{}{}
\SetKwFunction{GuessKey}{GuessKey}
\KwIn{Four pairs of special plain-cipher images $(\bm{I}_0, \bm {I}'_0)$, $(\bm{I}_{85}, \bm{I}'_{85})$, $(\bm{I}_{170}, \bm {I}'_{170}), (\bm{I}_{255}, \bm{I}'_{255})$, and the recovered equivalent sub-key $\bm{S}_3$}
\KwOut{Candidate equivalent sub-keys $E_1$, $D_1$, and $\bm{S}_4$}
\Fn{\GuessKey{$\bm{I}, \bm{I}', \bm{S}_3, s, t$}}{
	$\bm{I}_r\leftarrow$ obtain Red plane of $\bm{I}$\;
	$\bm{I}'_r\leftarrow$ obtain Red plane of $\bm{I}'$\;
	$\bm {I}^{\star \star}_r \oplus \bm S_4 ~\leftarrow$ via Eq.~(\ref{eq:cipher_r_decry}) and $\bm{I}'_r$\;
	$\hat{\bm{I}}^{\star\star}_r\leftarrow$ encrypt $\bm{I}_r$ with $f_s(\cdot)$, $g(\cdot)$, $\bm{S}_3$, and $f_t^{-1}(\cdot)$\;
	$\hat{\bm{S}}_4\leftarrow$ $\bm{I}^{\star \star}_{r} \oplus \bm S_4 \oplus \hat {\bm{I}}^{\star \star}_{r}$\;
	\Return {$\hat {\bm{S}}_4$}\;
}
\For{$s\leftarrow 0$ \KwTo $7$}{
	\For{$t\leftarrow 0$ \KwTo $7$}{
		$\hat {\bm{S}}_{4, 0} \leftarrow$ \GuessKey{$\bm{I}_0, \bm{I}'_0, \bm{S}_3, s, t$}\;
		$\hat {\bm{S}}_{4, 85} \leftarrow$ \GuessKey{$\bm{I}_{85}, \bm{I}'_{85}, \bm{S}_3, s, t$}\;
		$\hat {\bm{S}}_{4, 170} \leftarrow$ \GuessKey{$\bm{I}_{170}, \bm{I}'_{170}, \bm{S}_3, s, t$}\;
		$\hat {\bm{S}}_{4, 255} \leftarrow$ \GuessKey{$\bm{I}_{255}, \bm{I}'_{255}, \bm{S}_3, s, t$}\;
		\If {$\hat{\bm{S}}_{4, 0}==\hat{\bm{S}}_{4, 85}==\hat{\bm{S}}_{4, 170}==\hat{\bm{S}}_{4,255}$}{
			$E_1 \leftarrow s$\;
			$D_1 \leftarrow t$\;
			$\bm{S}_4\leftarrow \hat{\bm{S}}_{4, 0}$\;
			Print candidate sub-keys $E_1$, $D_1$, and $\bm{S}_4$\;
		}
	}
}
\end{algorithm}

\begin{savenotes}
	\begin{table*}[!htb]
		\centering
		\begin{threeparttable}
			\caption{Four classes of candidate values of $(E_1, D_1, \bm{S}_4)$.}
			\begin{tabular}{c *{3}{|c}}
				\hline
				Equivalent sub-key class  $\mathcal{K}_{r1}$\footnote{$\bm{\hat{S}}_4^{(i, j)}$ represents $j^{th}$ different keystream $\bm{S}_4$ in the $i^{th}$ equivalent classes, where
$i\in\{1, 2, 3, 4\}, j\in\{1, 2, \cdots, 8\}$. Because the value of $\bm{\hat{S}}_4^{(i,j)}$ depends on the original sub-key $(Y_{{4}}(0), \mu_4)$, only the more general relationships between them are given here.}
				 & Equivalent sub-key class $\mathcal{K}_{r2}$   &Equivalent sub-key class $\mathcal{K}_{r3}$  &Equivalent sub-key class $\mathcal{K}_{r4}$ \\ \hline
				$(0, 0, \bm{\hat{S}}_4^{(1, 1)})$   $,~~(4,4,\bm{\hat{S}}_4^{(1, 1)})$
				&$(0, 1, \bm{\hat{S}}_4^{(2, 1)}),~~(4, 5, \bm{\hat{S}}_4^{(2, 1)})$
				&$(1, 1, \bm{\hat{S}}_4^{(3, 1)}),~~(5, 5, \bm{\hat{S}}_4^{(3, 1)})$   &$(1, 0, \bm{\hat{S}}_4^{(4, 1)}),~~(5, 4, \bm{\hat{S}}_4^{(4, 1)})$ \\
				
				$(0, 2, \bm{\hat{S}}_4^{(1, 2)}),~~(4, 7, \bm{\hat{S}}_4^{(1, 2)})$    &$(0, 3, \bm{\hat{S}}_4^{(2, 2)}),~~(4, 6, \bm{\hat{S}}_4^{(2, 2)})$
				&$(1, 3, \bm{\hat{S}}_4^{(3, 2)}),~~(5, 6, \bm{\hat{S}}_4^{(3, 2)})$   &$(1, 2, \bm{\hat{S}}_4^{(4, 2)}),~~(5, 7, \bm{\hat{S}}_4^{(4, 2)})$ \\
				
				$(0, 4, \bm{\hat{S}}_4^{(1, 3)}),~~(4, 0, \bm{\hat{S}}_4^{(1, 3)})$    &$(0, 5, \bm{\hat{S}}_4^{(2, 3)}),~~(4, 1, \bm{\hat{S}}_4^{(2, 3)})$
				&$(1, 5, \bm{\hat{S}}_4^{(3, 3)}),~~(5, 1, \bm{\hat{S}}_4^{(3, 3)})$   &$(1, 4, \bm{\hat{S}}_4^{(4, 3)}),~~(5, 0, \bm{\hat{S}}_4^{(4, 3)})$ \\
				
				$(0, 7, \bm{\hat{S}}_4^{(1, 4)}),~~(4, 2, \bm{\hat{S}}_4^{(1, 4)})$    &$(0, 6, \bm{\hat{S}}_4^{(2, 4)}),~~(4, 3, \bm{\hat{S}}_4^{(1, 4)})$
				&$(1, 6, \bm{\hat{S}}_4^{(3, 4)}),~~(5, 3, \bm{\hat{S}}_4^{(3, 4)})$   &$(1, 7, \bm{\hat{S}}_4^{(4, 4)}),~~(5, 2, \bm{\hat{S}}_4^{(4, 4)})$ \\
				
				$(2, 0, \bm{\hat{S}}_4^{(1, 5)}),~~(7, 4, \bm{\hat{S}}_4^{(1, 5)})$    &$(2, 1, \bm{\hat{S}}_4^{(2, 5)}),~~(7, 5, \bm{\hat{S}}_4^{(2, 5)})$
				&$(3, 1, \bm{\hat{S}}_4^{(3, 5)}),~~(6, 5, \bm{\hat{S}}_4^{(3, 5)})$   &$(3, 0, \bm{\hat{S}}_4^{(4, 5)}),~~(6, 4, \bm{\hat{S}}_4^{(4, 5)})$ \\
				
				$(2, 2, \bm{\hat{S}}_4^{(1, 6)}),~~(7, 7, \bm{\hat{S}}_4^{(1, 6)})$    &$(2, 3, \bm{\hat{S}}_4^{(2, 6)}),~~(7, 6, \bm{\hat{S}}_4^{(2, 6)})$
				&$(3, 3, \bm{\hat{S}}_4^{(3, 6)}),~~(6, 6, \bm{\hat{S}}_4^{(3, 6)})$   &$(3, 2, \bm{\hat{S}}_4^{(4, 6)}),~~(6, 7, \bm{\hat{S}}_4^{(4, 6)})$ \\
				
				$(2, 4, \bm{\hat{S}}_4^{(1, 7)}),~~(7, 0, \bm{\hat{S}}_4^{(1, 7)})$    &$(2, 5, \bm{\hat{S}}_4^{(2, 7)}),~~(7, 1, \bm{\hat{S}}_4^{(2, 7)})$
				&$(3, 5, \bm{\hat{S}}_4^{(3, 7)}),~~(6, 1, \bm{\hat{S}}_4^{(3, 7)})$   &$(3, 4, \bm{\hat{S}}_4^{(4, 7)}),~~(6, 0, \bm{\hat{S}}_4^{(4, 7)})$ \\
				
				$(2, 7, \bm{\hat{S}}_4^{(1, 8)}),~~(7, 2, \bm{\hat{S}}_4^{(1, 8)})$    &$(2, 6, \bm{\hat{S}}_4^{(2, 8)}),~~(7, 3, \bm{\hat{S}}_4^{(2, 8)})$
				&$(3, 6, \bm{\hat{S}}_4^{(3, 8)}),~~(6, 3, \bm{\hat{S}}_4^{(3, 8)})$   &$(3, 7, \bm{\hat{S}}_4^{(4, 8)}),~~(6, 2, \bm{\hat{S}}_4^{(4, 8)})$ \\ \hline
			\end{tabular}
            \label{tab:equivalent_keys_classes}
		\end{threeparttable}
	\end{table*}
\end{savenotes}

Multiple groups of candidate sub-keys $(E_1$, $D_1, \bm{S}_4)$ are outputted by running Algorithm~\ref{alg:keysearch}.
It is further found that any original sub-key $(E_1$, $D_1, \bm{S}_4)$ has 16 sets of possible results via a number of experiments
(for more detail see Table~\ref{tab:equivalent_keys_classes}).

The underlying reasons can be divided into the following two cases:
\begin{itemize}	
\item Different DNA encoding and decoding rules may have the same DNA mapping and encryption effect.
Referring to Property~\ref{prop:dna_mapping_num}, although there are 64 combinations of $s$ and $t$,
$F_{s, t}(\cdot)$ and $G_{s, t}(\cdot)$ have only 8 and 16 different cases, respectively.
Consequently, different parameter pairs $(s_1, t_1)$ and $(s_2, t_2)$ still generate the same transformation effect for the DNA mapping.
Furthermore, if the corresponding keystream $\bm{S}_4$ has the same value, then the corresponding encryption results are the same.
Such cases can be found in Table~\ref{tab:equivalent_keys_classes}, such as $(0, 0, \bm{\hat{S}}_4^{(1, 1)})$ and $(4, 4, \bm{\hat{S}}_4^{(1, 1)})$.

\item Different DNA mappings may have the same encryption effect.
From Properties~\ref{prop:fx_xor_keys} and~\ref{prop:gx_xor_keys}, there are two different DNA mappings $F_{s, t}(\cdot)$ that generate the same encryption effect.
So does $G_{s, t}(\cdot)$. Assuming that one CEK-R (the Red channel) is $(s_1, t_1, \bm{S}_4)$, there is another one $(s_2, t_2, \bm{S}_4\oplus \bm{\Lambda})$,
where $\bm{\Lambda}=\{\lambda, \lambda, \cdots, \lambda\}$. Such cases can also be found in Table~\ref{tab:equivalent_keys_classes}; that is,
$(0, 0, \bm{\hat{S}}_4^{(1, 1)})$ and $(0, 2, \bm{\hat{S}}_4^{(1, 2)})$.
\end{itemize}

Note that because the three color channels are not completely independent (See Eq.~(\ref{eq:dna_diffusion})),
the 16 CEK-Rs that are obtained here are not the final attacking result.

\subsubsection{Cracking encryption operations in the Green channel and the Blue channel}

This subsection discusses how to attack the encryption operations in the Green channel and the Blue channel, which
are also used to confirm the attacking results for the Red channel further.

First, from Eq.~(\ref{eq:pixel_diffusion}), one has
\begin{equation}
\label{eq:cipher_gb_decry}
	\left\{
	\begin{split}
 I^{\star \star}_b(i)\oplus  \mathit{DS}(i)              & = I'_{\rm t}(i), \\
 I^{\star \star}_g(i)\oplus S_5(i) \oplus \mathit{DS}(i) & = I'_g(i)\oplus{I}'_g(i-1)\oplus I'_{\rm t}(i),
	\end{split}
	\right.
\end{equation}
where $\mathit{DS}(i)=S_4(i)\oplus S_6(i)$, $I'_{\rm t}(i)={I}'_b(i)\oplus{I}'_b(i-1)\oplus {I}'_r(i)\oplus{I}'_r(i-1)$.
This suggests that one can decrypt the data in the two channels of cipher-image $\bm{I}'_b$ and $\bm{I}'_g$ into the intermediate images
$\bm{I}^{\star \star}_b\oplus \bm S_6\oplus\bm S_4$, and $\bm{I}^{\star \star}_g\oplus\bm S_5\oplus\bm S_6\oplus\bm S_4$, respectively.
As for the attacking in the Green channel, a similar search method as Algorithm~\ref{alg:keysearch} is performed with the same four plain-images and the corresponding cipher-images.
However, one point is different: the case in the Green channel is not completely independent. As shown in Eqs.~(\ref{eq:2bits}), (\ref{eq:dna_trans}), (\ref{eq:dna_gx}), and (\ref{eq:dna_diffusion}), this is related to DNA encoding (sub-key $E_1$) and DNA complement (key $\bm{S}_3$) of the red channel.
So, it needs to exhaustively search $8\times 8\times 8= 512$ times for the Green channel.
Finally, one can obtain some candidate equivalent sub-keys for the Green channel (CEK-Gs), $(E_1, E_2, D_2, \bm{S}_4\oplus\bm{S}_6)$.

As for the blue channel, the attacking method is also very similar. The size of search space here is $8\times 8\times 8=512$.
The output is some candidate equivalent sub-keys for the blue channel (CEK-Bs) $(E_2, E_3, D_3, \bm{S}_5 \oplus \bm{S}_6 \oplus \bm{S}_4)$.

Among these three output results, CEK-Rs, CEK-Gs, and CEK-Bs, $E_1$ or $E_2$ appear twice, but CEK-Rs and CEK-Gs are not exactly the same,
and so do CEK-Gs and CEK-Bs. If an element appears in the two candidate sets at the same time, then it is an optional key and reserved; otherwise, it is a wrong key and removed.
An illustrating example can be found in Sec.~\ref{sec:attack example}.

\subsubsection{Revealing the permutation sub-key}
\label{sec:reveal_permu_key}

Up to now, the equivalent sub-keys shown in Table~\ref{tab:equ_keys} are all known except for the permutation index $\bm{S}_1$.
So, one can recover the permuted image $\{\bm{I}^{\star}_r, \bm{I}^{\star}_g, \bm{I}^{\star}_b\}$ from a cipher-image $\{\bm{I}'_r, \bm{I}'_g, \bm{I}'_b\}$.
Additionally, the corresponding plain-image $\{\bm{I}_r, \bm{I}_g, \bm{I}_b\}$ is accessible for the attacker under the scenario of a chosen-plaintext attack.
In this process, MPPS is degenerated to a position permutation-only encryption scheme, whose security performance has been studied in depth in \cite{Lcq:Optimal:SP11,Cqli:Scramble:IM17}.
Using the sub-optimal attacking method based on multi-branch tree proposed in \cite{Lcq:Optimal:SP11}, it only needs to select $\lceil\log_{256}(3L)\rceil$ special plain-images to
attack the position permutation part of MPPS, where function $\lceil x \rceil$ returns the smallest possible integer value which is greater than or equal to the argument $x$.
In all, the whole equivalent secret key of MPPS can be successfully recovered with only $\lceil\log_{256}(3L)\rceil+4$ chosen plain-images and the corresponding cipher-images.

\begin{table*}[!htb]
\resizebox{\textwidth}{!}{
	\begin{threeparttable}
		\centering
		\caption{All possible equivalent keys for the three color channels.}
		\begin{tabular}{c *{2}{|c}}
			\hline
			$(E_1, D_1, \bm{\hat{S}}_4)$ for R channel &$(E_1, E_2, D_2, \bm{\hat{S}}_5)$ for G channel &$(E_2, E_3, D_3, \bm{\hat{S}}_6 )$ for B channel\\ \hline			
			$ {({\color{red}{0}},1,\bm{\hat{S}}_4^{(1)})}, {({\color{red} {4}},5,\bm{\hat{S}}_4^{(1)})}$
			&$ {({\color{red} {0,1}}, 0, \bm{\hat{S}}_5^{(1)})}, {({\color{red} {0,5}},4,\bm{\hat{S}}_5^{(1)})}, {({\color{red} {4,1}},4,\bm{\hat{S}}_5^{(1)})}, {({\color{red} {4, 5}}, 0, \bm{\hat{S}}_5^{(1)})}$
			&$ {({\color{red} {1}}, 1, 1, \bm{\hat{S}}_6^{(1)})}, {({\color{red} {1}},5,5,\bm{\hat{S}}_6^{(1)})}, {({\color{red} {5}},1,5,\bm{\hat{S}}_6^{(1)})}, {({\color{red} {5}}, 5, 1, \bm{\hat{S}}_6^{(1)})}$ \\

			$ {({\color{red}{0}}, 3, \bm{\hat{S}}_4^{(2)})}, {({\color{red} {4}},6,\bm{\hat{S}}_4^{(2)})}$
			&$ {({\color{red}{0,1}}, 2, \bm{\hat{S}}_5^{(2)})}, {({\color{red} {0,5}},7,\bm{\hat{S}}_5^{(2)})}, {({\color{red} {4,1}},7,\bm{\hat{S}}_5^{(2)})}, {({\color{red} {4,5}},2,\bm{\hat{S}}_5^{(2)})}$
			&$ {({\color{red}{1}}, 1, 3, \bm{\hat{S}}_6^{(2)})}, {({\color{red} {1}},5,6,\bm{\hat{S}}_6^{(2)})}, {({\color{red} {5}},1,6,\bm{\hat{S}}_6^{(2)})}, {({\color{red} {5}},5,3,\bm{\hat{S}}_6^{(2)})}$ \\

			$ {({\color{red} {0}},5,\bm{\hat{S}}_4^{(3)})}, {({\color{red} {4}},1,\bm{\hat{S}}_4^{(3)})}$
			&$ {({\color{red} {0,1}},4,\bm{\hat{S}}_5^{(3)})}, {({\color{red} {0,5}},0,\bm{\hat{S}}_5^{(3)})},  {({\color{red} {4,1}},0,\bm{\hat{S}}_5^{(3)})}, {({\color{red} {4,5}},4,\bm{\hat{S}}_5^{(3)})}$
			&$ {({\color{red} {1}},1,5,\bm{\hat{S}}_6^{(3)})}, {({\color{red} {1}},5,1,\bm{\hat{S}}_6^{(3)})},  {({\color{red} {5}},1,1,\bm{\hat{S}}_6^{(3)})}, {({\color{red} {5}},5,5,\bm{\hat{S}}_6^{(3)})}$ \\

			$ {({\color{red} {0}},6,\bm{\hat{S}}_4^{(4)})}, {({\color{red} {4}},3,\bm{\hat{S}}_4^{(4)})}$
			&$ {({\color{red} {0, 1}},7,\bm{\hat{S}}_5^{(4)})}, {({\color{red} {0,5}},2,\bm{\hat{S}}_5^{(4)})},  {({\color{red} {4,1}},2,\bm{\hat{S}}_5^{(4)})}, {({\color{red} {4,5}},7,\bm{\hat{S}}_5^{(4)})}$
			&$ {({\color{red} {1}},1,6,\bm{\hat{S}}_6^{(4)})}, {({\color{red} {1}},5,3,\bm{\hat{S}}_6^{(4)})},  {({\color{red} {5}},1,3,\bm{\hat{S}}_6^{(4)})}, {({\color{red} {5}},5,6,\bm{\hat{S}}_6^{(4)})}$ \\

			$ {(2,1,\bm{\hat{S}}_4^{(5)})},{(7,5,\bm{\hat{S}}_4^{(5)})}$
			& $ {(1, 0, 0, \bm{\hat{S}}_5^{(5)})}, {(1,4,4,\bm{\hat{S}}_5^{(5)})},  {(5,0,4,\bm{\hat{S}}_5^{(5)})}, {(5, 4, 0,\bm{\hat{S}}_5^{(5)})}$
			& $ {(3, 3, 1, \bm{\hat{S}}_6^{(5)})}, {(3,6,5,\bm{\hat{S}}_6^{(5)})},  {(6,3,5,\bm{\hat{S}}_6^{(5)})}, {(6, 6, 1,\bm{\hat{S}}_6^{(5)})}$ \\

			${(2,3,\bm{\hat{S}}_4^{(6)})},{(7,6,\bm{\hat{S}}_4^{(6)})}$
			& $ {(1, 0, 2, \bm{\hat{S}}_5^{(6)})}, {(1,4,7,\bm{\hat{S}}_5^{(6)})},  {(5,0,7,\bm{\hat{S}}_5^{(6)})}, {(5, 4, 2,\bm{\hat{S}}_5^{(6)})}$
			& $ {(3, 3, 3, \bm{\hat{S}}_6^{(6)})}, {(3,6,6,\bm{\hat{S}}_6^{(6)})},  {(6,3,6,\bm{\hat{S}}_6^{(6)})}, {(6, 6, 3,\bm{\hat{S}}_6^{(6)})}$ \\

			${(2, 5, \bm{\hat{S}}_4^{(7)})},{(7,1,\bm{\hat{S}}_4^{(7)})}$
			&$ {(1, 0, 4, \bm{\hat{S}}_5^{(7)})},{(1,4,0,\bm{\hat{S}}_5^{(7)})}, {(5,0,0,\bm{\hat{S}}_5^{(7)})}, {(5, 4, 4, \bm{\hat{S}}_5^{(7)})}$
			&$ {(3, 3, 5,\bm{\hat{S}}_6^{(7)})},{(3,6,1,\bm{\hat{S}}_6^{(7)})}, {(6,3,1,\bm{\hat{S}}_6^{(7)})}, {(6, 6, 5, \bm{\hat{S}}_6^{(7)})}$  \\

			${(2, 6, \bm{\hat{S}}_4^{(8)})},{(7,3,\bm{\hat{S}}_4^{(8)})}$
			&${(1, 0, 7, \bm{\hat{S}}_5^{(8)})},{(1,4,2,\bm{\hat{S}}_5^{(8)})}, {(5,0,2,\bm{\hat{S}}_5^{(8)})}, {(5, 4, 7, \bm{\hat{S}}_5^{(8)})}$
			&${(3, 3, 6, \bm{\hat{S}}_6^{(8)})},{(3,6,3,\bm{\hat{S}}_6^{(8)})}, {(6,3,3,\bm{\hat{S}}_6^{(8)})}, {(6, 6, 6, \bm{\hat{S}}_6^{(8)})}$  \\
         \hline
		\end{tabular}
    \label{tab:reveal_equivalent_keys}		
	\end{threeparttable}
}
\end{table*}

\begin{figure*}[!htb]
	\centering
	\includegraphics[scale=0.49]{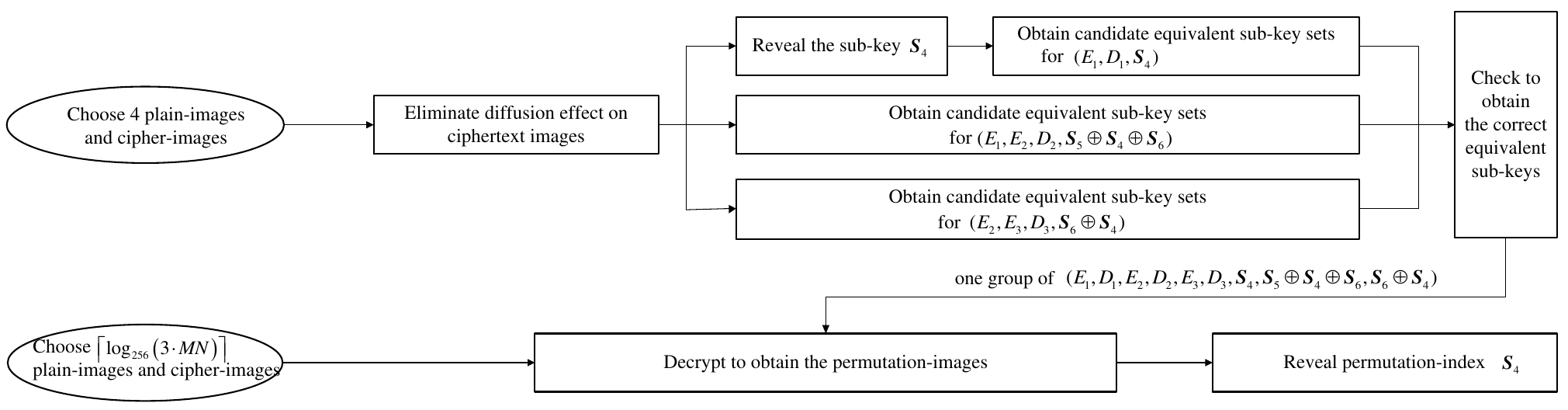}
	\caption{The framework for obtaining all equivalent sub-keys.}
	\label{fig:attack_process}
\end{figure*}

\subsection{Attacking process and simulation results}
\label{sec:attack example}

\begin{table}[!htb]
	\centering
	\caption{The result of attacking MPPS with a chosen-plaintext attack.}
	\begin{tabular}{ll}
		\hline
	 	Original sub-keys   & The obtained equivalent sub-keys\\ \hline
		$K_1=(0.11, 3.91)$  & $\bm{S}_1 = \{9, 1, 10, 2, 11, 5, 7, 3, 12, 6, 4, 8\}$\\
		$K_2=(0.12, 3.92)$  & $\bm{S}_2 = \{4, 5, 5, 1, 7, 6\}$\\
		$K_3=(0.13, 3.93)$ 	& $\bm{S}_3 = \{0, 0, 1, 1, 1, 1, , 0, 1, 1, 0, 1, 0, 1, 1, 0\}$\\
		$K_4=(0.14, 3.94)$  & $\bm{S}_4 = \{99, 172, 189, 130\}$\\
		$K_5=(0.15, 3.95)$	& $\bm{S}_5 = \{155, 45, 47, 189\}$\\
		$K_6=(0.16, 3.96)$	& $\bm{S}_6 = \{193, 122, 164, 238\}$ \\ \hline		
	\end{tabular}
\label{tab:keys_setup}
\end{table}

To verify the performance of the proposed chosen-plaintext attack, a large number of experiments were performed with some random secret keys.
To facilitate an illustration with the limited presentation size, the RGB colour images of size $2\times 2 \times 3$ (i.e., $M=N=2$) are used.
In the following simulation, the initial iterations of CLS map and CTS map are both set as $t_1=t_2=500$, and the six original sub-keys of MPPS are as
listed in the second column of Table~\ref{tab:keys_setup}.
Accordingly, the equivalent sub-key corresponding to each original key is calculated by the \textit{initialization} procedure of MPPS.

As shown in Fig.~\ref{fig:attack_process}, two groups of special plain-images were constructed to obtain all of the equivalent keys of MPPS.
The first group of four special plain-images and the corresponding cipher-images are given here:
\begin{equation}
\label{eq:firstgroup}
\notag
\begin{split}
&\bm {I}_0      = \{ 0, 0, \cdots,   0\},\\
&\bm {I}'_0     = \{198,  60, 216, 204; 107,  69, 102,  24;  49,  72, 224, 205\};\\
&\bm {I}_{255}  = \{255,  255, \cdots, 255\};\\
&\bm {I}'_{255} = \{ 51, 158,  39, 228; 148,  69, 153,  24; 110, 234, 181, 229\};\\
&\bm {I}_{85}   = \{ 85, 85, \cdots, 85\},\\
&\bm {I}'_{85}  = \{108,  60, 114,  204; 59,  20,  51,  12;  49,  72, 224, 205\};\\
&\bm {I}_{170}  = \{170, 170, \cdots, 170\},\\
&\bm {I}'_{170} = \{153, 158, 141, 228; 145, 20, 153,  12; 110, 234, 181, 229\}.
\end{split}
\end{equation}
First, referring to Eqs.~(\ref{eq:cipher_r_decry}),~(\ref{eq:cipher_gb_decry}),
the above cipher-images can be partially decrypt without keys, that is,
\begin{equation}
\label{eq:new_cipher_image}
\notag
\begin{split}
&\bm {I}^{**}_0     =  \{198, 250, 228,  20;  156, 173, 111,  71; 247, 131,  76,  57\};\\
&\bm {I}^{**}_{255} =  \{ 51, 173, 185,  195; 201, 248,  58,  18;  93,  41, 230, 147\};\\
&\bm {I}^{**}_{85}  =  \{108,  80,  78,  190; 102,   6, 193, 172;  93,  41, 230, 147\};\\
&\bm {I}^{**}_{170} =  \{153,   7,  19,  105; 102,   6, 193, 172; 247, 131,  76,  57\},
\end{split}
\end{equation}
where $\bm{I}^{**}_{\phi}=\{\bm{I}^{**}_{r \phi}, \bm {I}^{**}_{g \phi}, \bm {I}^{**}_{b \phi}\}=\{\bm{I}^{\star\star}_{r \phi} \oplus \bm S_4,
\bm{I}^{\star \star}_{g \phi}\oplus \bm S_5 \oplus \bm S_{6} \oplus \bm S_4, \bm{I}^{\star\star}_{b\phi} \oplus \bm S_{6} \oplus \bm S_4 \}$,
and $\phi\in\{0, 85, 170, 255\}$.
Then, calculate the difference
$\Delta \bm {I}^{**}_{r} = \bm {I}^{**}_{r0} \oplus \bm {I}^{**}_{r255}=\bm {I}^{\star \star}_{r0} \oplus \bm {I}^{\star \star}_{r255} = \Delta \bm {I}^{\star \star}_{r} = \{(11110101)_2, (01010111)_2, (01011101)_2, (11010111)_2 \}$.
By using Eq.~(\ref{eq:reveal_s3}), one can determine the binary sequence
\begin{equation}
\label{eq:example_reveal_S3}
\bm{S}_3 = \{0, 0, 1, 1, 1, 1, 1, 0, 1, 1, 0, 1, 0, 1, 1, 0\}.
\end{equation}
By adopting Algorithm~\ref{alg:keysearch} with $\bm{S}_3$ and the aforementioned four pairs of plain-images and the corresponding cipher-images,
one can obtain a candidate for the equivalent sub-key set of $(E_1, D_1, \bm{\hat{S}}_4)$.
Similarly, the counterparts for the Green and Blue channels can be produced;
namely, the candidates for the equivalent sub-key sets of $(E_1, E_2, D_2, \bm{\hat{S}}_5)$ and $(E_2, E_3, D_3, \bm{\hat{S}}_6)$,
where $\bm{\hat{S}}_4 = \bm{{S}}_4, \bm{\hat{S}}_5=\bm{{S}}_5 \oplus \bm{{S}}_4 \oplus \bm{{S}}_6$ and
$\bm{\hat{S}}_6 = \bm{S}_6\oplus\bm{S}_4$.
Let $\bm{\hat{S}}_m^{(j)}$ represent the $j^{th}$ different keystream $\bm{\hat{S}}_m$, $j\in\{1, 2, \dots, 8\}$, $m\in \{4, 5, 6\}$.
Table~\ref{tab:reveal_equivalent_keys} shows the whole test results obtained by running the attack algorithms on the three channels separately,
where the red number signals that $E_i$ appears in both columns.
However, they need to be confirmed further by checking the consistency.
For example, one can get $E_1 \not\in\{2, 7, 1, 5\}$ as they do not appear in the first two columns.
By excluding the results, one can obtain the correct equivalent sub-key set of MPPS,
which is marked in red in Table~\ref{tab:reveal_equivalent_keys}.
Meanwhile, another group of images are chosen to obtain the permutation index $\bm{S}_1$,
which includes only one (i.e. $ \lceil\log_{256}(3 \times 2 \times 2) \rceil=1$) plain-image chosen.
For example, one can select the following special plain-image and obtain the corresponding cipher-image:
\begin{equation*}
\begin{split}
 \bm {I}_n  &= \{  0,   1,   2,   3;   4,   5,   6,   7;   8,   9,   10, 11\},\\
 \bm {I}'_n & = \{202,  48, 210, 196;  99,  72, 106,   17;  60, 76,  231, 205\}.
\end{split}
\end{equation*}

An attacker can decrypt cipher-image $\bm{I}'_n$ step-by-step with a decryption key,
which is selected from the correct sets in Table~\ref{tab:reveal_equivalent_keys}, such as $(E_1, D_1, \bm{\hat{S}}_4)= (0, 1, \{201, 6, 23, 40\})$,
$(E_1, E_2, D_2, \bm{\hat{S}}_5)= (0, 1, 0, \{147, 81, 156, 123\})$, and $(E_2, E_3$, $D_3, \bm{\hat{S}}_6)$ $= (1, 1, 1, \{247, 131, 76, 57\})$.
The concrete decryption process includes the following two steps:
\begin{enumerate}[label=\alph*)]
\item By using the inverse of Eqs.~(\ref{eq:cipher_r_decry}), (\ref{eq:cipher_gb_decry}) with the decryption sub-keys $\bm{\hat{S}}_4$, $\bm{\hat{S}}_5$, $\bm{\hat{S}}_6$,
one can decrypt cipher-image $\bm{I}'_n$ into an image $\bm{I}^{**}_n=\{169, 95, 86, 148; 172, 93, 90, 150; 84, 80, 92, 80\}$.

\item By adopting DNA coding, subtraction (the inverse of Eq.~(\ref{eq:dna_diffusion})), complement (the inverse of Eqs.~(\ref{eq:dna_trans}), (\ref{eq:dna_gx}), and DNA decoding with the decryption sub-keys
$(E_1, D_1, E_2, D_2, E_3, D_3, \bm{S}_3)$, one can get a scrambled image $\bm{I}^{{\star}}_n=\{8, 0, 9, 1, 10, 4, 6$, $2, 11, 5, 3, 7\}$.
\end{enumerate}

\begin{figure*}[!htb]
\centering
\begin{minipage}{\BigOneImW}
	\centering
	\includegraphics[width=\BigOneImW]{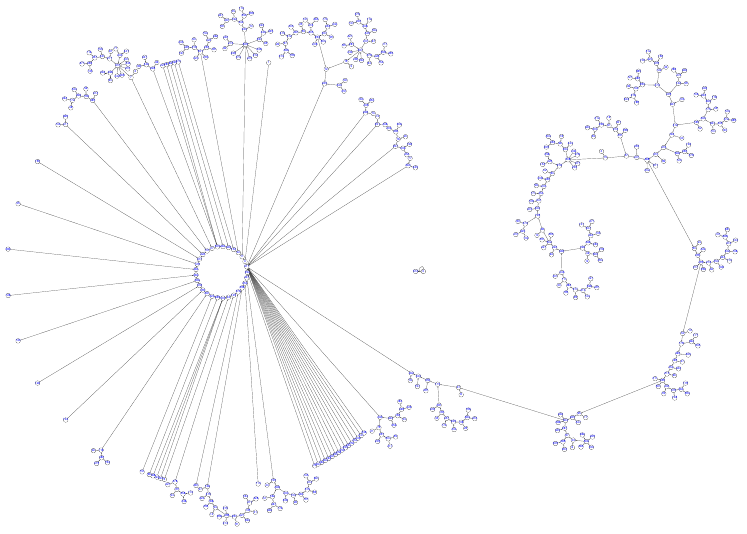}
	a)
\end{minipage}\hspace{\figsep}
\begin{minipage}{\BigOneImW}
	\centering
	\includegraphics[width=\BigOneImW]{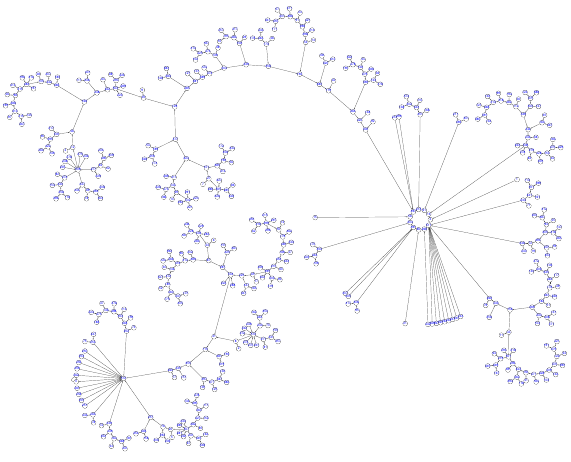}
	b)
\end{minipage}\hspace{\figsep}
\begin{minipage}{\BigOneImW}
	\centering
	\includegraphics[width=\BigOneImW]{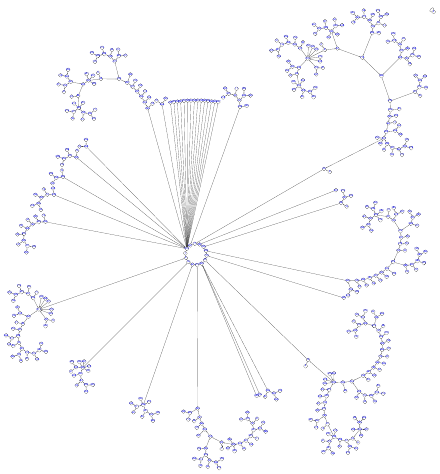}
	c)
\end{minipage}
\caption{The functional graph of CLS map ($\mu=121/2^5$) with 9-bit fixed-point precision and different quantization strategies:
a) floor; b) round; c) ceiling, where number $i$ in each node represents value $i/2^9$.}
\label{fig:networkCLS9bits}
\end{figure*}

Finally, by comparing the position of every pixel in $\bm{I}_n$ with that of the same gray value in $\bm {I}^{\star}_n$,
one can determine a permutation index $\bm{S}_1=\{9, 1, 10, 2, 11, 5, 7, 3, 12, 6, 4, 8\}.$
In all, by selecting $num=4+\lceil \log_{256}(3 \times 2 \times 2)\rceil=5$ plain-images, one can first uniquely determine the equivalent sub-keys $\bm{S}_1$ and $\bm{S}_3$,
which agree with Table~\ref{tab:keys_setup}. In addition, there are eight possible values of $(E_1, D_1, \bm{\hat{S}}_4)$ in the table.
If $(E_1, D_1, \bm{\hat{S}}_4)$ is one of them (fixed value), then there are eight possible values of $(E_2, D_2, \bm{\hat{S}}_5)$.
Similarly, there are eight possible values of $(E_3, D_3, \bm{\hat{S}}_6)$ when $(E_2, D_2, \bm{\hat{S}}_5)$ is a fixed value.
Therefore, there are $8\times 8\times 8=512$ sets of equivalent secret keys in total and one of them exists in Table~\ref{tab:keys_setup}.

\begin{figure*}[!htb]
\centering
\begin{minipage}{\FourImW}
\centering
\includegraphics[width=\FourImW]{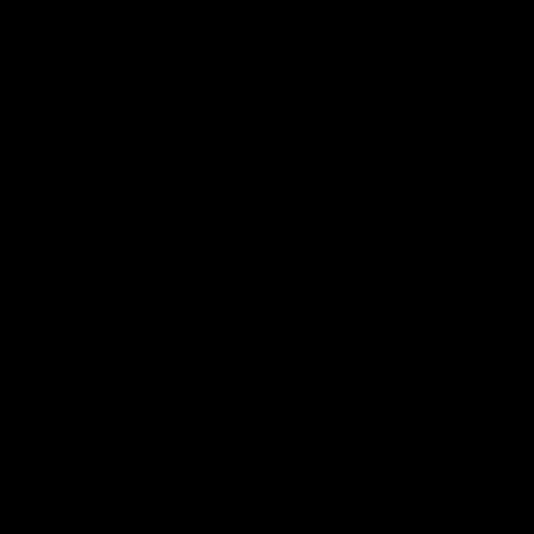}
a1)
\end{minipage}\hspace{\figsep}
\begin{minipage}{\FourImW}
\centering
\includegraphics[width=\FourImW]{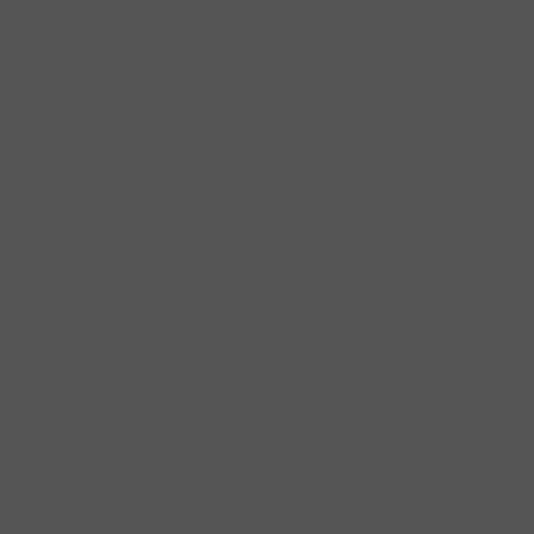}
b1)
\end{minipage}\hspace{\figsep}
\begin{minipage}{\FourImW}
\centering
\includegraphics[width=\FourImW]{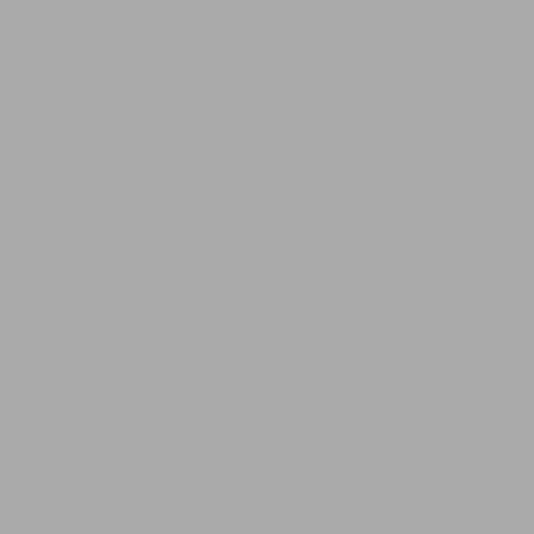}
c1)
\end{minipage}\hspace{\figsep}
\begin{minipage}{\FourImW}
\centering
\includegraphics[width=\FourImW]{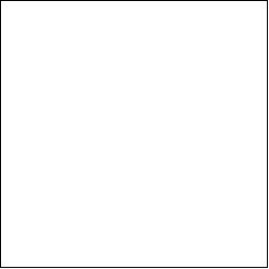}
d1)
\end{minipage}\\
\begin{minipage}{\FourImW}
\centering
\includegraphics[width=\FourImW]{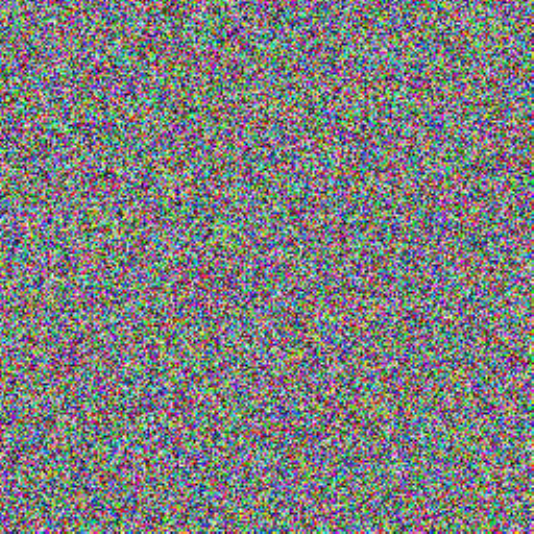}
a2)
\end{minipage}\hspace{\figsep}
\begin{minipage}{\FourImW}
\centering
\includegraphics[width=\FourImW]{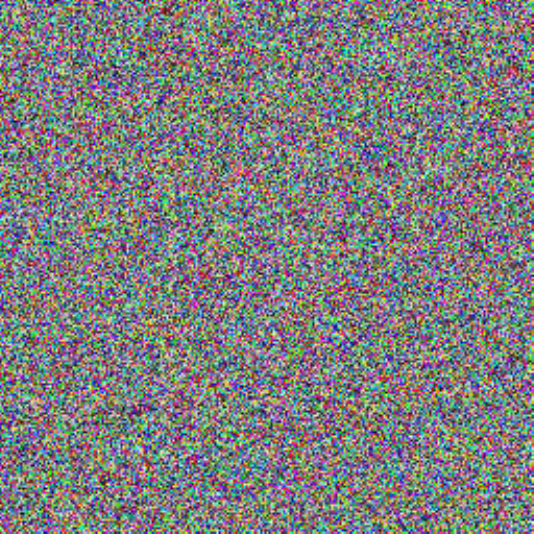}
b2)
\end{minipage}\hspace{\figsep}
\begin{minipage}{\FourImW}
\centering
\includegraphics[width=\FourImW]{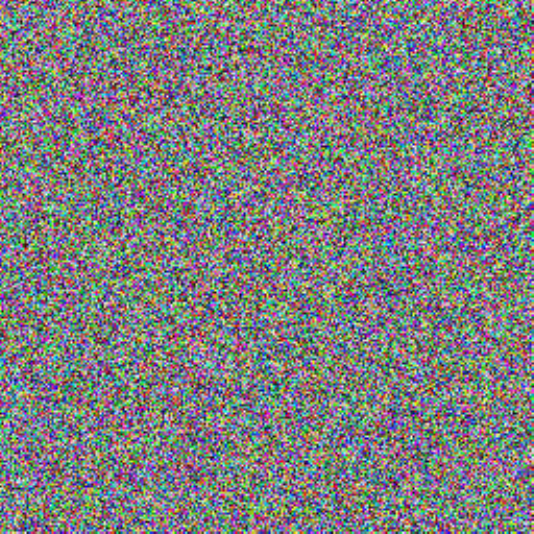}
c2)
\end{minipage}\hspace{\figsep}
\begin{minipage}{\FourImW}
\centering
\includegraphics[width=\FourImW]{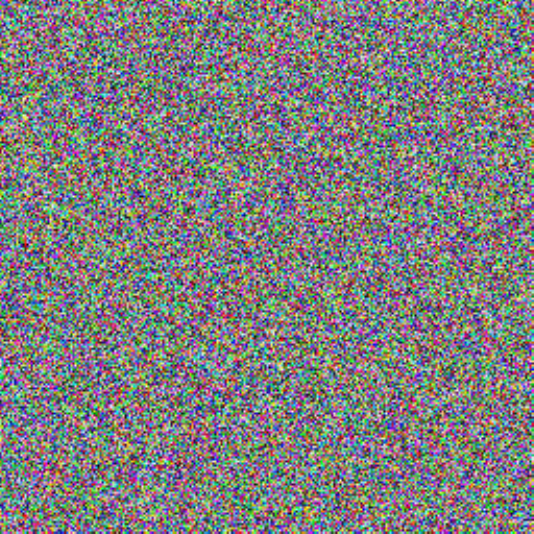}
d2)
\end{minipage}
\caption{Four pairs of special plain-cipher images in the proposed attack:
a1-d1) four plain-images $\bm{I}_0$, $\bm{I}_{85}$, $\bm{I}_{170}$ and $\bm {I}_{255}$ of size $256\times 256$;
a2-d2) the corresponding cipher-images $\bm{I}'_0$, $\bm{I}'_{85}$, $\bm {I}'_{170}$, and $\bm{I}'_{255}$.}
\label{fig:special_image_pairs}
\end{figure*}

\begin{figure*}[!htb]
\centering
\begin{minipage}{0.84\linewidth}
\centering
\includegraphics[width=0.84\linewidth]{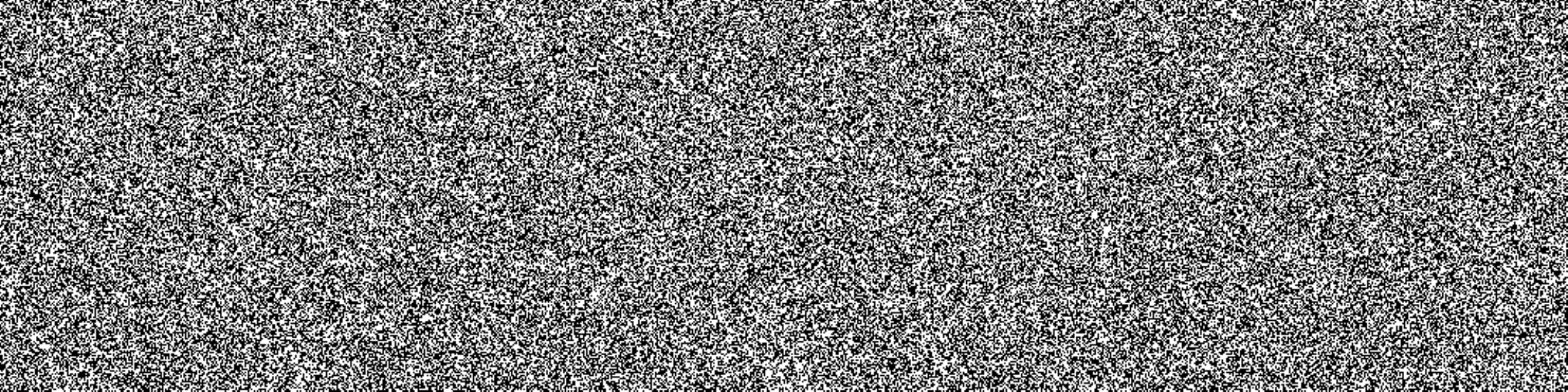}
\\a)
\end{minipage}
\begin{minipage}{\FourImW}
\centering
\includegraphics[width=\FourImW]{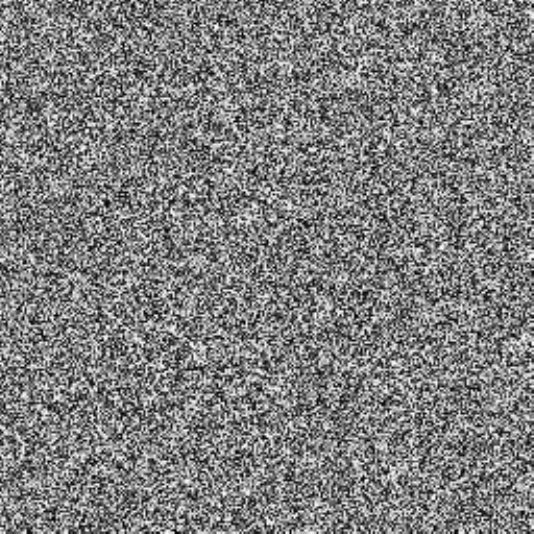}
b)
\end{minipage}\hspace{\figsep}
\begin{minipage}{\FourImW}
\centering
\includegraphics[width=\FourImW]{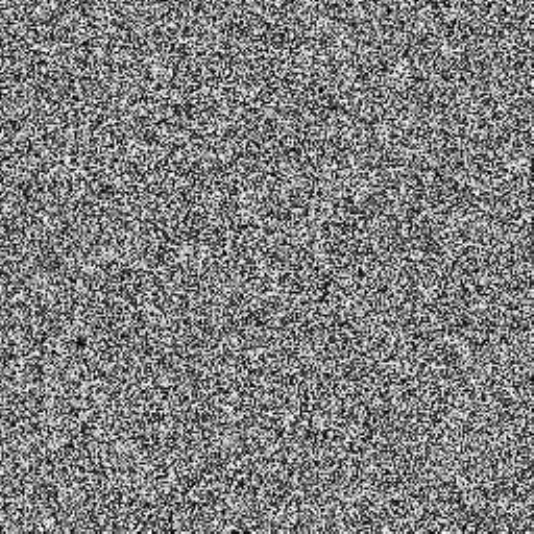}
c)
\end{minipage}\hspace{\figsep}
\begin{minipage}{\FourImW}
\centering
\includegraphics[width=\FourImW]{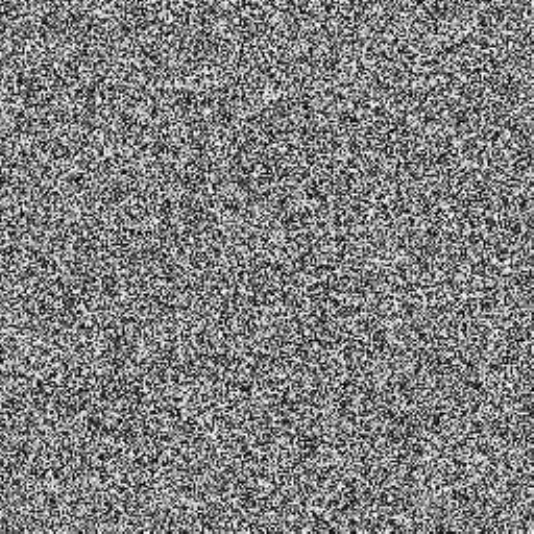}
d)
\end{minipage}
\caption{The recovered equivalent sub-keys: a) the sub-key $\bm{S_3}$ of size $256 \times 1024$;
		b-d) the sub-keys $\bm{\hat{S}}_4$, $\bm{\hat{S}}_5$ and $\bm{\hat{S}}_6$ of size $256 \times 256$.}
\label{fig:keystream_image}
\end{figure*}

\begin{figure*}[!htb]
\centering
\begin{minipage}{\FourImW}
\centering
\includegraphics[width=\FourImW]{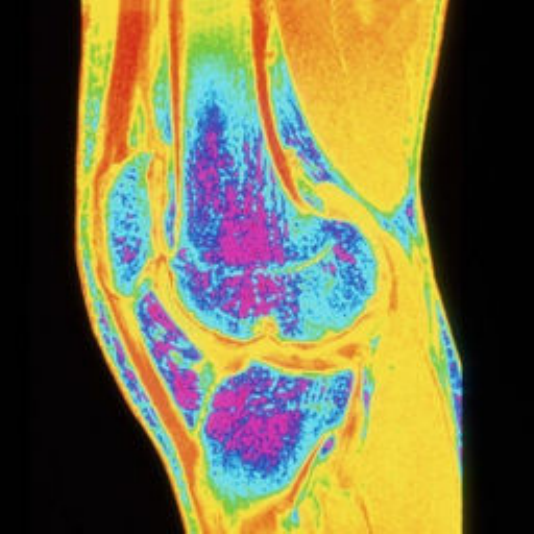}
a1)
\end{minipage}\hspace{\figsep}
\begin{minipage}{\FourImW}
\centering
\includegraphics[width=\FourImW]{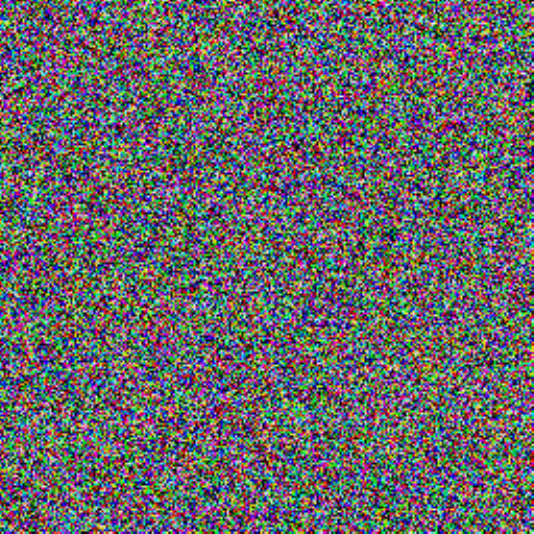}
b1)
\end{minipage}\hspace{\figsep}
\begin{minipage}{\FourImW}
\centering
\includegraphics[width=\FourImW]{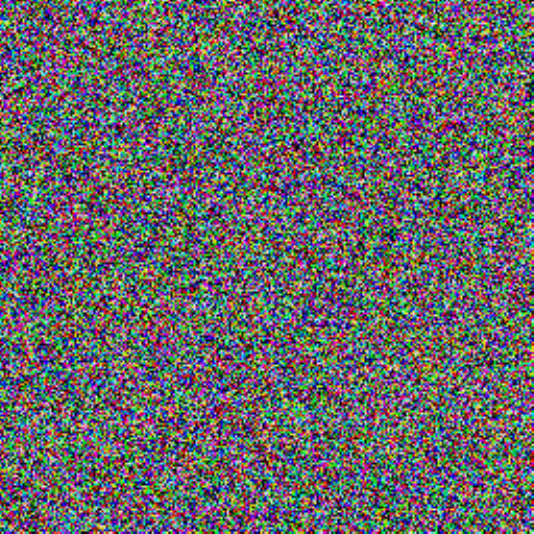}
c1)
\end{minipage}\hspace{\figsep}
\begin{minipage}{\FourImW}
\centering
\includegraphics[width=\FourImW]{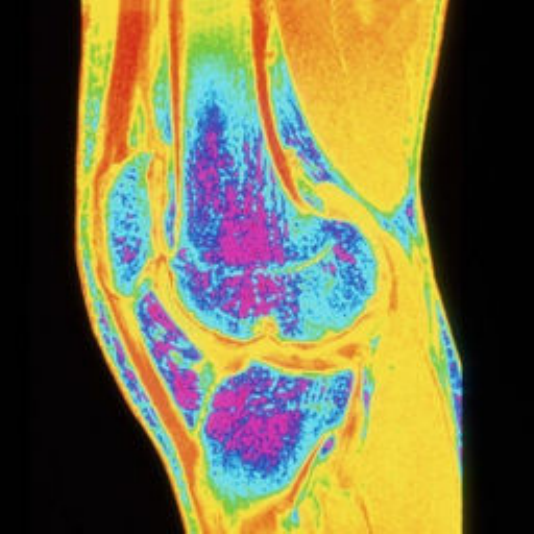}
d1)
\end{minipage}\\
\begin{minipage}{\FourImW}
\centering
\includegraphics[width=\FourImW]{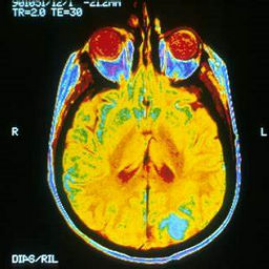}
a2)
\end{minipage}\hspace{\figsep}
\begin{minipage}{\FourImW}
\centering
\includegraphics[width=\FourImW]{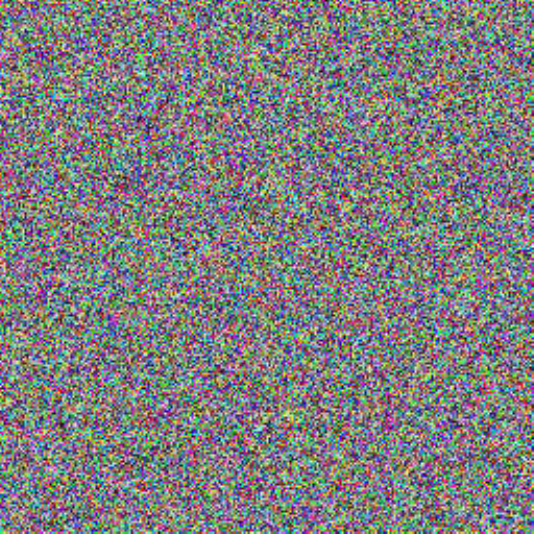}
b2)
\end{minipage}\hspace{\figsep}
\begin{minipage}{\FourImW}
\centering
\includegraphics[width=\FourImW]{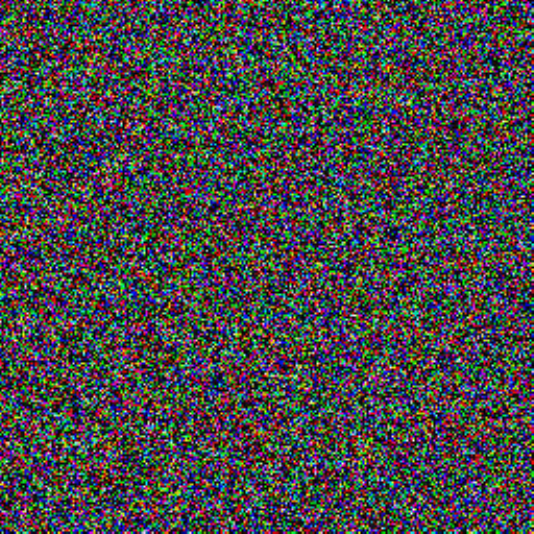}
c2)
\end{minipage}\hspace{\figsep}	
\begin{minipage}{\FourImW}
\centering
\includegraphics[width=\FourImW]{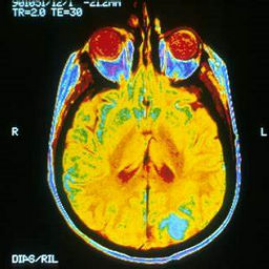}
d2)
\end{minipage}
\caption{Simulation results of the proposed attack on two encrypted images:
		a1-a2) two original-images $\bm{I}$ of size $256\times 256$ unknown to the attacker;
		b1-b2) the corresponding encrypted images $\bm {I}'$ available to the attacker;
		c1-c2) the recovered scrambled image $\bm{I}^{{\star}}_n$ from $\bm{I}'$ with the sub-keys
		$\bm{\hat{S}}_4$, $\bm{\hat{S}}_5$, $\bm{\hat{S}}_6$, $\bm{S_3}$, and $(E_1, E_2, E_3, D_1, D_2, D_3)=(0, 1, 1, 0, 1, 1)$;
		d1-d2) the decrypted plain-images from $\bm{I}^{{\star}}_n$ with the key $\bm{S}_1$.}.
\label{fig:decipher_process}
\end{figure*}

To further demonstrate the effectiveness of the proposed attack, a number of simulations on some RGB colour image of size $256\times 256$ are conducted with the same original key that was used in the previous example.
Referring to the attack framework in Fig.~\ref{fig:attack_process}, the equivalent sub-keys $\bm {S_2} = ({E}_1,{E}_2,{E}_3, {D}_1,{D}_2,{D}_3) = (0,1,1,0,1,1)$,  $\bm {S}_3$, $\bm{\hat{S}}_4$, $\bm{\hat{S}}_5$ and $\bm{\hat{S}}_6$ can be obtained according to four special plain-cipher image pairs $(\bm {I}_0,\bm {I}'_0)$, $(\bm {I}_{85},\bm {I}'_{85})$, $(\bm {I}_{170},\bm {I}'_{170})$ and $(\bm {I}_{255},\bm {I}'_{255})$, which are described in Fig.~\ref{fig:special_image_pairs} and Fig.~\ref{fig:keystream_image}.
Then, selecting $\lceil \log_{256}(3 \times 256 \times 256) \rceil = 3$ special images by the attack method mentioned in Sec.~\ref{sec:reveal_permu_key} to reveal the equivalent permutation key $\bm {S_1}$.
After obtaining all of these sub-keys, one can recover any encrypted image.
Figure~\ref{fig:decipher_process} shows the process of decrypting two cipher-images,
which indicates that the proposed attacks are effective.

\begin{figure*}[!htb]
\centering
\begin{minipage}{\BigOneImW}
\centering
\includegraphics[width=\BigOneImW]{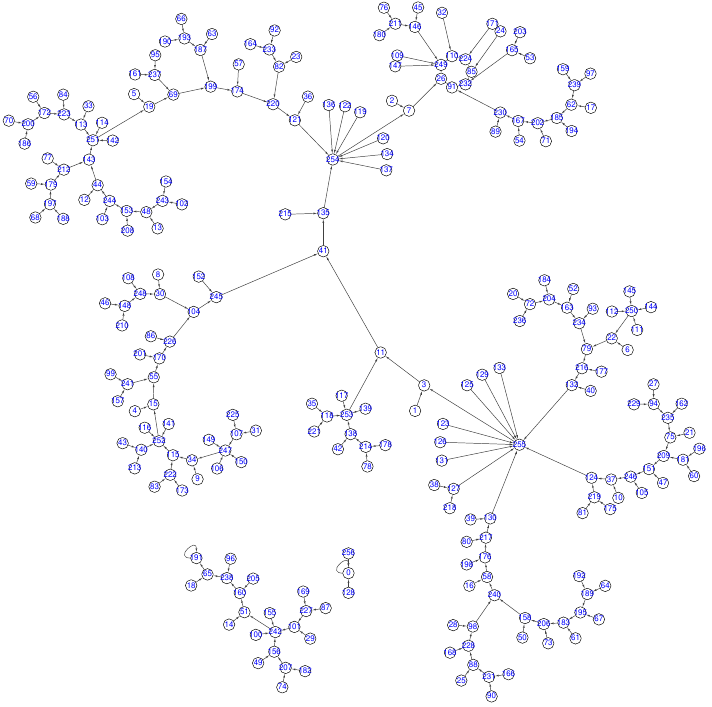}
a)
\end{minipage}\hspace{\figsep}
\begin{minipage}{\BigOneImW}
\centering
\includegraphics[width=\BigOneImW]{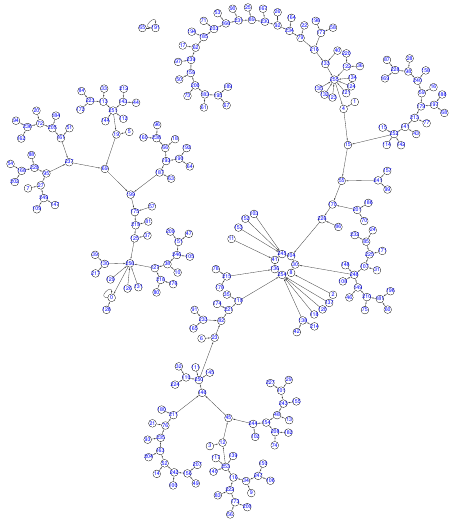}
b)
\end{minipage}\hspace{\figsep}
\begin{minipage}{\BigOneImW}
\centering
\includegraphics[width=\BigOneImW]{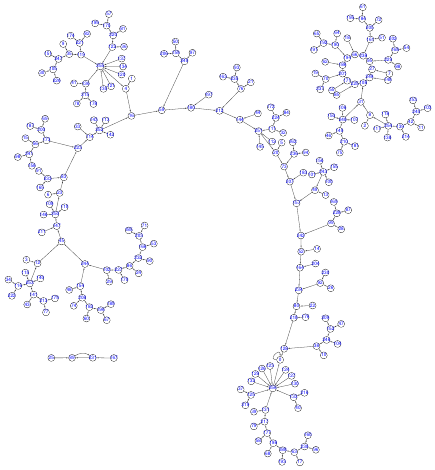}
c)
\end{minipage}
\caption{A functional graph of the CLT map ($\mu=123/2^5$) with 8-bit fixed-point precision and different quantization strategies:
a) floor; b) round; c) ceiling, where the number $i$ in each node denotes value $i/2^8$.}
\label{fig:networkCLT6bitsfloorroundceil}
\end{figure*}

 \begin{figure}[!htb]
  \centering
  \begin{minipage}{\OneImW}
  \centering
  \includegraphics[width=\OneImW]{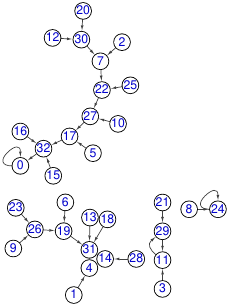}
    a)
  \end{minipage}
  \hspace{6pt}
  \begin{minipage}{\OneImW}
  \centering
  \includegraphics[width=\OneImW]{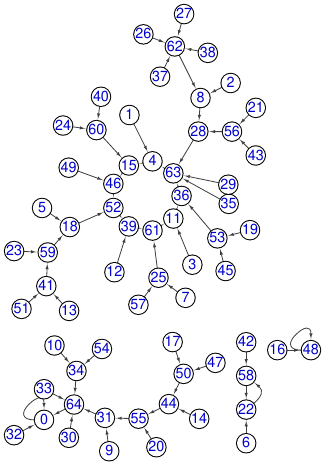}
    b)
  \end{minipage}\\
  \centering
  \begin{minipage}{\BigOneImW}
  \centering
  \includegraphics[width=\BigOneImW]{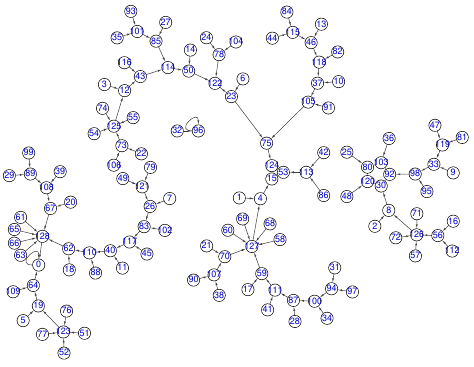}
    c)
  \end{minipage}
  \begin{minipage}{\BigOneImW}
  \centering
  \includegraphics[width=\BigOneImW]{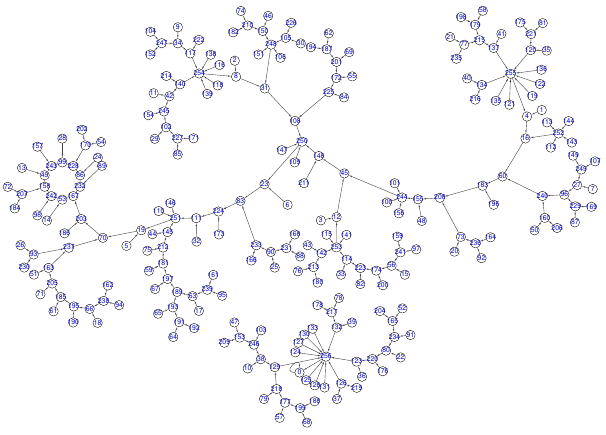}
    d)
  \end{minipage}
  \caption{A functional graph of the generalized CLS map ($\mu=61/2^4$) with $n$-bit fixed-point precision and round quantization:
a) $n=5$; b) $n=6$; c) $n=7$; d) $n=8$.}
  \label{fig:CLS5678}
\end{figure}

\begin{figure}[!htb]
\centering
\begin{minipage}{0.8\OneImW}
  \centering
  \includegraphics[width=0.8\OneImW]{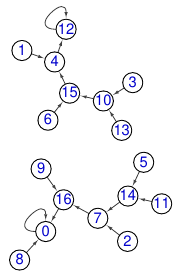}
    a)
  \end{minipage}
  \hspace{2pt}
  \begin{minipage}{\OneImW}
  \centering
  \includegraphics[width=\OneImW]{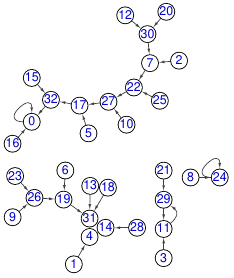}
    b)
  \end{minipage}\\
  \begin{minipage}{\BigOneImW}
  \centering
  \includegraphics[width=\BigOneImW]{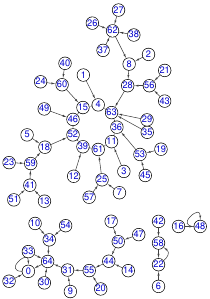}
    c)
  \end{minipage}
  \hspace{2pt}
  \begin{minipage}{1.2\BigOneImW}
  \centering
  \includegraphics[width=1.2\BigOneImW]{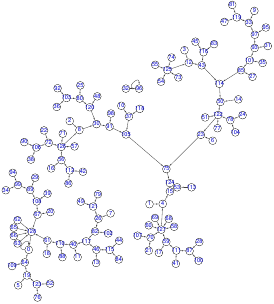}
    d)
\end{minipage}
\caption{A functional graph of the generalized CLT map ($\mu=63/2^4$) with $n$-bit fixed-point precision and round quantization:
a) $n=4$; b) $n=5$; c) $n=6$; d) $n=7$, where the number $i$ in each node denotes value $i/2^n$.}
\label{fig:CLT4567}
\end{figure}

\subsection{Evaluation of other attack analyses}

In \cite[Sec.~IV]{Ravichandran:DNA:ITN2017}, the performance of MPPS was analyzed from eight aspects.
Here, their credibilities are re-evaluated one by one as follows.
\begin{itemize}
\item \textit{Keyspace}: the statement ``it can be evident that the coupled chaotic map achieves amplified chaotic range in the entire region
of [0, 4]'' in \cite{Ravichandran:DNA:ITN2017} is questionable. The size of each sub-figure in \cite[Fig.~1]{Ravichandran:DNA:ITN2017} is about 0.9 inch by 0.9 inch.
Assume that the adopted  dots per inch (dpi) is 200, then the number of the printed dots is only about 32,400.
Observing \cite[Fig.~1]{Ravichandran:DNA:ITN2017}, one can see that the distribution is not uniform. Meanwhile, the effective keyspace of MPPS is far overestimated, due to the following four types of equivalent or weak keys in MPPS:
\begin{itemize}
\item \textit{Equivalent keys type I} (coupled chaotic maps): since the symmetry of the initial values of CLS~(Eq.~(\ref{eq:cls})) or CLT~ (Eq.~(\ref{eq:cts}));
that is, $f_{\textup{CLS}}(Y_i)=f_{\textup {CLS}}(1-Y_i)$ or $f_{\textup {CLT}}(Y_j)=f_{\textup{CLT}}(1-Y_j)$, where $i\in\{1, 4, 5, 6\}$ and $j\in\{2, 3\}$.
In this case, the effective keyspace of MPPS is reduced to $\frac{1}{64}$ (i.e., $\frac{1}{2^4\times 2^2}$) of the claimed size.

\item \textit{Equivalent keys type II} (DNA coding rules): given that there are three kinds of DNA encoding or decoding rules (corresponding sub-key $K_2$) in
three color channels, the effective keyspace of $(E_1, E_2, E_3, D_1, D_2, D_3)$ is only $8^6=2^{18}$ and is not $10^{28}$ ($10^{28} \approx 2^{93}$), as was claimed in \cite{Ravichandran:DNA:ITN2017}.
Considering DNA encoding properties (Property~\ref{prop:dna_relation}, Property~\ref{prop:dna_mapping_num}), there are many equivalent keys in Table~\ref{tab:reveal_equivalent_keys}, and the effective keyspace of $K_2$ is further reduced to $8^3=2^9$.

\item \textit{Weak keys type I} (coupled chaotic maps):
	 The initial values $Y(0) = 0,  1.0$ or $0.5$ are the fixed points of CCS or CLT; that is, $f_{\textup {CLS}}(0) = f_{\textup {CLT}}(0) =  0$, $f_{\textup {CLS}}(1.0) = f_{\textup {CLT}}(1.0) = 0$ and $f_{\textup {CLS}}(0.5) = f_{\textup {CLT}}(0.5) = 1.0$.
	 Then, one can conclude that there are about $(3\times (10^{28})^5) \times 6 = 2^{469}$ weak keys (i.e., ineffective keys) in this case.
	
\item \textit{Weak keys type II} (DNA coding rules):
When the DNA encoding and decoding rules are the same (i.e., $E_i=D_i=j$), the DNA transformations have no any encryption effect,
where $i\in\{1, 2, 3\}$ and $j\in\{0, 1, 2, 3, 4, 5, 6, 7\}$. Thus, MPPS has $(10^{28}\times \frac{8^{6}}{64^{6}}\times (10^{28})^5)=2^{540}$ weak keys.
\end{itemize}

The influence of the invalid key items on real key space are summarized in Table~\ref{tab:invalid_key},
where the keyspace of MPPS is estimated to be approximately $(2^{93})^6=2^{558}$.
As for the equivalent keys type II, the real key space of $K_2$ is $2^9$, so the whole cryptosystem is $2^9\times(2^{93})^5=\frac{2^{558}}{2^{84}}$.
Considering the influence of the four factors in Table~\ref{tab:invalid_key}, the actual key space can be estimated as $O(\frac{2^{558}}{2^{84}})=O(2^{474})$.

\item \textit{Key sensitivity}: As analyzed in \cite{Cqli:Scramble:IM17,cqli:Entropy:IEEEA18,Uhl:chaos:TIFS2018}, there are many equivalent secret keys due to the dynamics degradation of the chaotic system in a finite-precision computer. As shown in Fig.~\ref{fig:networkCLT6bitsfloorroundceil}, \ref{fig:networkCLS9bits}, one can see that many initial states evolve into the same orbit after some iterations of the involved map.
Similar to digitized Logistic map and Tent map analyzed in \cite{cqli:meet:JISA19}, the degradation patterns of CLS map and CLT map are similar (regardless of the implementation precision and quantization schemes).
Due to the presentation limitation, the SMNs of the two analyzed coupled maps under four small precision are given in Fig.~\ref{fig:CLS5678} and Fig.~\ref{fig:CLT4567}, respectively.
Recalling Sec.~\ref{algorithm}, the initial states are discarded to avoid the so-called ``transient effect", which makes the key insensitivity problem become more severe.

\item \textit{Statistical attack}: Statistical tests, such as histogram, entropy and correlation coefficient analysis can only provide a necessary but not sufficient test for security evaluation.
In other words, passing such tests cannot verify that the involved schemes are secure against various attacks \cite{cqli:Entropy:IEEEA18,Uhl:chaos:TIFS2018}.

As shown in \cite{Uhl:chaos:TIFS2018}, several counterexamples are constructed to demonstrate that an obviously insecure encryption scheme can still perform well or can pass the aforementioned tests.
In \cite{cqli:Entropy:IEEEA18}, a counterexample of image histogram analysis is given. These are enough to show that statistical tests cannot be used as the decisive criteria for evaluating the security performance of an encryption scheme.

\item \textit{Differential attack}: In \cite{Ravichandran:DNA:ITN2017}, it is claimed that ``the value of NPCR and UACI of the cipher images are nearer to the ideal value (NPCR $>99\%$) (UACI $\approx 33\%$)'',
thus ``this evidences the ability of suggested algorithm to tolerate differential attacks''. In fact, similar to the analysis of the statistical attack,  this conclusion is highly questionable because NPCR and UACI evaluations are essentially statistical analysis. In fact, the diffusion rules (Eq.~(\ref{eq:pixel_diffusion})) of MPPS are vulnerable to differential attack.
According to Property~\ref{prop:differ}, any difference between cipher-images is unrelated to the sub-keys $\bm{S}_4$, $\bm{S}_5$, and $\bm{S}_6$,
which means that the effective key length is reduced by half.

\item \textit{Chosen-plaintext attack}: One specific relationship between plain-image and the corresponding cipher-image is almost unrelated with the capability of an encryption scheme against a chosen-plaintext attack.
In \cite[Sec.~4.5]{Ravichandran:DNA:ITN2017}, the criteria to determine whether to resist a plaintext attack is invalid. As a counter-example, a permutation-only encryption scheme is obviously a very weak against
the chosen-plaintext attack, whereas the criteria provided by \cite[Sec.~4.5]{Ravichandran:DNA:ITN2017} leads to an opposite conclusion.

\item \textit{Cropping Attack}: If the encryption process is unrelated to the plain-image, then any encryption scheme is robust against
cropping attack. Actually, the ability against cropping attacks is contradictory to the good sensitivity of ciphertext on the change of plaintext.
In \cite[Fig.~8]{Ravichandran:DNA:ITN2017}, the difference between the two decrypted images is very close, which means the sensitivity of cipher-image on
change of plain-image is weak. In other words, these experiments show that MPPS has security flaws rather than superiority.

\item \textit{Performance comparison}: As emphasized in \cite{cqli:meet:JISA19}, any encryption scheme should target a specific application scenario; otherwise, the balancing point among usability, security and efficiency is obscure. Meanwhile, in \cite[Table~11]{Ravichandran:DNA:ITN2017}, most of the performance comparisons use statistical analysis,
which hardly supports the view that the proposed scheme is superior to the other similar chaos-based encryption schemes.
\end{itemize}

\begin{table*}[!htb]
	\centering
	\caption{Quantitative analysis of the influence of the invalid key items on real key space.}
	\begin{tabular}{ccccc}
		\hline\noalign{\smallskip}
		Invalid key items    & Equivalent keys type I
		& Equivalent keys type II
		& Weak keys type I
		& Weak keys type II\\
		\noalign{\smallskip}\hline\noalign{\smallskip}
		Effective keyspace size
		& $\frac{2^{558}}{2^{6}}$  &  $\frac{2^{558}}{2^{84}}$
		& $2^{558}-2^{469}$ &  $2^{558}-2^{540}$\\
		\noalign{\smallskip}\hline
	\end{tabular}
	\label{tab:invalid_key}       
\end{table*}

\subsection{Other defects}

\begin{itemize}
\item \textit{Practical consideration}: In \cite[Sec.~III]{Ravichandran:DNA:ITN2017}, it is stated that
``selection of a region of interest and region of non-interest is solely left to the user under the recommended physician''.
Besides, MPPS ``requires the user's feed that specifies the number of rounds of operation''.
Consequently, it requires professional training for ordinary users.

\item \textit{Efficiency}: As analyzed in \cite{cqli:Entropy:IEEEA18}, conversion functions (\ref{eq:convert14}), (\ref{eq:q_s4})
waste much computation on the discarded bits. To generate a sequence of length six $\bm{X}_2$,
$t_2=500$ states are wasted.
Moreover, it is questionable that DNA encoding and decoding can improve the efficiency of an encryption scheme.
As shown in Fig.~\ref{fig:dna_encryption_framework}, the DNA coding region (namely DNA encoding, complement, diffusion,
and decoding processes) of MPPS can implement large-scale parallel computing on all two bits of image pixels.
However, various DNA computation processes can be regarded as a series of S-box operations that are defined in domain $\mathbb{Z}_4$.
As summarized in Property~\ref{prop:dna_mapping_num}, the DNA encoding and decoding process in the Red plane can be seen as $(8+16)=24$ S-boxes,
whose selection is determined by the encoding and decoding rules (key $K_2$ ($\bm{S}_2$)) and a binary random number (key $K_3$ ($\bm{S}_3$));
that is, when $S_3=0$, $F_{s, t}: \mathbb{Z}_4 \rightarrow \mathbb{Z}_4$; otherwise, $G_{s, t}: \mathbb{Z}_4 \rightarrow \mathbb{Z}_4$.
In other words, DNA codec (bits-to-codes or codes-to-bits mapping) operations consume a large amount of time, as the authors claimed that
``the time complexity in pixel transformation and DNA encoding and decoding are $O(M \times N \times 3 \times 2)$''
and ``the time complexity for performing DNA addition is $O(12 \times M \times N)$'', but contribute nothing to the real security performance of MPPS.
\end{itemize}

\section{Conclusion}

This paper has re-analyzed the theoretical security and practical performance of a medical privacy protection scheme based on DNA encoding and chaotic maps.
Some interesting properties of DNA encoding are found and the scheme was rigorously proven to be insecure against the chosen-plaintext attack.
Detailed experimental results were provided to show more security defects, including the existence of a large number of weak secret keys,
weak key sensitivity, low efficiency, and bad usability.
The DNA-based encryption scheme that was analyzed is very important for promoting interdisciplinary research on application of DNA computing in cryptography, as follows:
a) Build a precise mathematical model for DNA-related operations;
b) Disclose the intrinsic relationship between DNA encoding and cryptographic primitives;
c) Establish a general cryptoanalyzing framework of the DNA-based encryption algorithms.
On the other hand, to resist the proposed attacks, some security enhancement mechanisms of such schemes are worthy of future research, such as adding nonlinear modules (e.g., S-box), or designing DNA encoding and decoding modules without key control.

\section*{Acknowledgement}

This work was supported by the National Natural Science Foundation of China (no. 61772447), and the China Postdoctoral Science Foundation
(no. 2019M660511).

\bibliographystyle{model5-names}
\bibliography{DNA_secu}

\end{document}